\documentclass[fleqn,usenatbib]{mnras}

\usepackage{newtxtext,newtxmath}
\usepackage[T1]{fontenc}

\DeclareRobustCommand{\VAN}[3]{#2}
\let\VANthebibliography\thebibliography
\def\thebibliography{\DeclareRobustCommand{\VAN}[3]{##3}\VANthebibliography}

\usepackage{graphicx}	% Including figure files
\usepackage{amsmath}	% Advanced maths commands

\newcommand{\lSect}[1]{{\label{sec:#1}}}
\newcommand{\lFig}[1]{{\label{fig:#1}}}

\newcommand{\FIGFF}[2]{{\ref{fig:#2}{#1}}}

\newcommand{\FIG}[2]{{Fig.~\FIGFF{#1}{#2}}}
\newcommand{\Fig}[1]{{\FIG{}{#1}}}

\newcommand{\Sectff}[1]{{\ref{sec:#1}}}
\newcommand{\Sect}[1]{{\S\Sectff{#1}}}

\newcommand{\Msun}{\ensuremath{\mathrm{M}_\odot}}

\newcommand{\KEPLER}{\ensuremath{\mathrm{\texttt{KEPLER}}}}
\newcommand{\MESA}{\ensuremath{\mathrm{\texttt{MESA}}}}
\newcommand{\mco}{\ensuremath{M_\mathrm{CO}}}
\newcommand{\mhe}{\ensuremath{M_\mathrm{He}}}
\newcommand{\xc}{\ensuremath{X_\mathrm{C}}}
\newcommand{\mfin}{\ensuremath{M_\mathrm{fin}}}

\title[Compact Object Distribution Comparison]{Comparing Compact Object Distributions from Mass- and Presupernova Core Structure-based Prescriptions}

\author[Patton et al.]{
Rachel A. Patton $^{1}$\thanks{E-mail: patton.502@osu.edu},
Tuguldur Sukhbold$^{1,2}$\thanks{NASA Hubble Fellow},
and J.J. Eldridge$^{3}$
\\
$^{1}$Department of Astronomy, The Ohio State University, 140 West 18th Ave, Columbus, OH 43210, USA\\
$^{2}$Center for Cosmology and AstroParticle Physics, The Ohio State University,
191 West Woodruff Avenue, Columbus, OH 43210\\
$^{3}$Department of Physics, University of Auckland, Private Bag 92019, Auckland, New Zealand\\
}

\date{Accepted XXX. Received YYY; in original form ZZZ}

\pubyear{2021}

\begin{document}
\label{firstpage}
\pagerange{\pageref{firstpage}--\pageref{lastpage}}
\maketitle

\begin{abstract}
Binary population synthesis (BPS) employs prescriptions to predict final fates, explosion or implosion, and remnant masses based on one or two stellar parameters at the evolutionary cutoff imposed by the code, usually at or near central carbon ignition. In doing this, BPS disregards the integral role late-stage evolution plays in determining the final fate, remnant type, and remnant mass within the neutrino-driven explosion paradigm. To highlight differences between a popular prescription which relies only on the core and final stellar mass and emerging methods which rely on a star's presupernova core structure, we generate a series of compact object distributions using three different methods for a sample population of single and binary stars computed in BPASS. The first method estimates remnant mass based on a star's carbon-oxygen (CO) core mass and final total mass. The second method uses the presupernova core structure based on the bare CO-core models of \citet{Pat20} combined with a parameterized explosion criterion to first determine final fate and remnant type, then remnant mass. The third method associates presupernova helium-core masses with remnant masses determined from public explosion models which rely implicitly on core structure. We find that the core-/final mass-based prescription favors lower mass remnants, including a large population of mass gap black holes, and predicts neutron star masses which span a wide range, whereas the structure-based prescriptions favor slightly higher mass remnants, mass gap black holes only as low as 3.5 \Msun, and predict neutron star mass distributions which cluster in a narrow range.  

\end{abstract}


%%%%%%%%%%%%%%%%%%%%%%%%%%%%%%%%%%%%%%%%%%%%%%%%%%

%%%%%%%%%%%%%%%%% BODY OF PAPER %%%%%%%%%%%%%%%%%%

\section{Introduction}
To understand the properties of how massive stars die and the compact objects they leave behind, we must understand how they live. Any prediction of the occurrence rate and formation channels of phenomena involving neutron stars (NSs) and stellar mass black holes (BHs) requires knowledge of the underlying birth distribution of such compact objects and the relative rates of the events (supernovae or implosions) which produce them.  Because nearly all massive (>8 \Msun) stars live in binaries or higher order systems \citep[e.g.,][]{San12,San13,Moe17}, the effects on these distributions caused by interaction with a companion must be considered since it can complicate a star’s evolution, final fate, and associated compact object \citep[e.g.,][]{DeM17,Sch21}.

Binary population synthesis (BPS) is often employed to capture these effects, such as mass transfer, common envelope evolution, and merging, for large, quasi-realistic populations of stars. High mass stars are particularly challenging to model because the physics governing their evolution, especially beyond core carbon ignition, remains uncertain and computationally expensive. To mitigate these uncertainties and limit the computational expense, BPS codes terminate a given massive star’s evolution at or near carbon ignition in the core. The commonly used ``rapid'' type of BPS codes do not actively evolve their stars at all, relying instead on the semi-analytic approximations of \citet{Hur00} and \citet{Hur02} that only approximately track a few global parameters until the asymptotic giant branch. 

Once evolution ceases, a final fate is assumed based on either initial or final properties of the star. This includes separating explosions from implosions, thus NSs and BHs, through an arbitrary value of initial mass \citep{Sra19}, helium core mass \mhe~\citep{Zap19}, simple energy arguments \citep{Eld04}, and carbon-oxygen (CO) core mass \mco~and final stellar mass \mfin~\citep[][hereafter F12]{Fry12}, the most commonly used prescription. The primary shortcoming of these simplistic approaches is their failure to capture the effects late-stage evolution has on a star’s core structure, and ultimately, its demise. An extensive body of work shows that a star’s final fate, within the neutrino-powered explosion paradigm, depends sensitively on the structure of a star’s core immediately preceding the iron core--collapse, which in turn is set by the last few thousand years of evolution in the stellar core \citep[e.g.,][]{Woo02,OCo11,Ugl12,Pej15,Suk16,Suk20}. Neglecting these stages leads to final fate and compact object predictions based on properties that may not strongly correlate with a star’s actual explodability.

Recently \citet[][(hereafter PS20]{Pat20} proposed a method to base a BPS prescription for final fates on a star’s pre-collapse core structure. The structure is based on bare CO-cores, whose evolutionary tracks are set by their starting mass and composition, evolved from central carbon ignition to core--collapse, filling in the evolutionary stages missed by the BPS code. The user can then infer a final structure for any star given their \mco\ and its starting composition, and apply their preferred structure-based criteria to reflect modern results on neutrino-driven explosion simulations. Since the PS20 approach requires knowledge of the initial CO-core composition, something which is not tracked in rapid BPS codes, currently it is only applicable to BPS calculations that are based on full stellar models.

However, there is an emerging alternative method, which we refer to as ``hybrid'', that applies results from detailed neutrino-driven explosion calculations to BPS simply through the terminal \mhe. This approach utilizes the correlation pointed out by \citet{Ert20} between the presupernova \mhe~and the properties of core--collapse, which allows BPS codes to determine the final properties based implicitly on the presupernova core structure without actually computing this structure (unlike in PS20). Recently \citet{Woo20} compiled a table directly connecting \mhe~to the compact remnant masses based on the calibrated neutrino-driven explosion models of \citet{Ert20} and pulsational pair-instability models from \citet{Woo19}, and used it to infer the birth mass distributions of NSs and BHs in various simple populations. \citet{Rom20} adopted a similar method, instead compiling the table based on the results from \citet{Suk16}, to predict BH-NS merger rates from a detailed population of full binary stars and comparing them to those predicted from F12. Since the presupernova \mhe~is well determined by the time of central carbon ignition, this simple and efficient method can be directly applied in any rapid BPS model.

To understand how the results based on these different prescriptions vary from one another it is important to apply them to the same sample population of stars. In this study, we apply the F12, PS20, and hybrid prescriptions to a suite of BPASS \citep[][]{Eld08,Eld09,Eld17,Sta18} single and binary models, comparing the distribution of compact objects determined from each prescription. In \Sect{mod} we discuss the BPASS models, the differences between the F12 and PS20 prescriptions, and how we construct the distributions. \Sect{res} compares the F12 and PS20 distributions. In \Sect{hybrid} we examine how the hybrid and PS20 distributions compare and we address the impact of the differences between the three prescriptions in \Sect{dis}. Finally, our results are summarized in \Sect{Con}.

\section{Models}
\lSect{mod}

\subsection{BPASS models}\lSect{BPASS}
We construct our distributions from a set of BPASS (v2.2) models, 10$^6$ \Msun\ of single stars and 10$^6$ \Msun\ of binary systems. BPASS is a hybrid binary population synthesis code which evolves its single stars and the primaries of binary stars with a detailed stellar evolution code while approximating the evolution of the secondaries using the semi-analytic evolution equations from \citet{Hur00,Hur02}, then evolving the secondaries in detail once the primary star’s evolution terminates. BPASS implements prescriptions for handling winds, mass transfer, common envelope evolution, mergers, orbital decay, and ultimately, the final fates of massive stars, and its underlying stellar models are based on those of \citet{Egg71} and \citet{Pol95}. 

Version 2.2 is largely equivalent to version 2.1 \citep[][]{Eld08,Eld09,Eld17} but has greater mass resolution between single star models than before. Binary models have coarser mass resolution than the single models. More importantly, BPASS v2.2 incorporates the updated binary birth parameter distributions of \citet[][see their Table 13]{Moe17}. BPS codes typically assume that a binary’s initial conditions, the mass of the primary ($M_1$), the mass ratio ($q=M_1/M_2$) between the two components, and the orbital period ($P$), are independent, with $q$ and $P$ having flat distributions and primary mass being drawn from an assumed initial mass function (IMF). However, there are strong observed trends in both $q$ and $P$ with respect to primary mass, which BPASS now considers. 

A detailed description of the initial model grid is given in \citet{Sta18}. Our set of single BPASS models span 0.2-100 \Msun~in steps of 0.1 \Msun~and our binary models span 4.5 – 100 \Msun\ for $M_1$ with increments of 0.5 \Msun. For the binaries, each primary star cycles through each possible combination of mass ratio and orbital period, with $q$ ranging between 0.1 and 0.9 in increments of 0.1, and $\log P$ ranging between 1.4 and 4 days in steps of 0.2 dex. We ignore binary models with log P below 1.4 to eliminate all merging systems, and to build a uniform coverage over the mass space we interpolate between models for each combination of $q$ and $\log P$. The twin star systems ($q = 1$) in BPASS are handled by incorporating its statistics from \citet{Moe17} into the systems with $q = 0.9$. All models are at solar ($Z = 0.02$) metallicity. Stars are evolved until core carbon ignition, where we extract the total mass of the star $M_\mathrm{fin}$, helium and CO- core masses $M_\mathrm{He}$, \mco, carbon mass fraction \xc, which we adopt as our proxy for composition, and the number of stars per million solar masses with a given set of initial conditions, as determined by an IMF and the \citet{Moe17} weights. BPASS assumes a \citet{Kro01} IMF for masses between 0.5 and 1 \Msun, and a \citet{Sal55} IMF for stars with masses heavier than 1 \Msun.

A few minor corrections were manually applied to the BPASS grid before the interpolation. Six single star models had anomalously low values of \mco~and/or \xc. We believe these were errors with the models and replaced them with \mco~and \xc~values linearly interpolated over the set of single star models. There were 25 entries missing from the binary models and an additional two binary models with anomalously low \xc, again, numerical errors with those specific models. Each missing or spurious model was replaced by an average of the models of the same initial orbital period and mass with mass ratio 0.1 higher and lower than the missing value. If the missing model had $q = 0.1$, we averaged the model with $q = 0.2$ of the same mass and orbital period with the $q = 0.1$ model of the same mass with the next lowest orbital period. If the missing model had $q = 0.9$, we averaged the model with $q = 0.8$ of the same mass and orbital period with the $q = 0.9$ models of the same mass and next highest orbital period.

\begin{figure}
    \centering
    \includegraphics[scale=0.54,trim={0.5cm 0.1cm 1.6cm 1.4cm},clip]{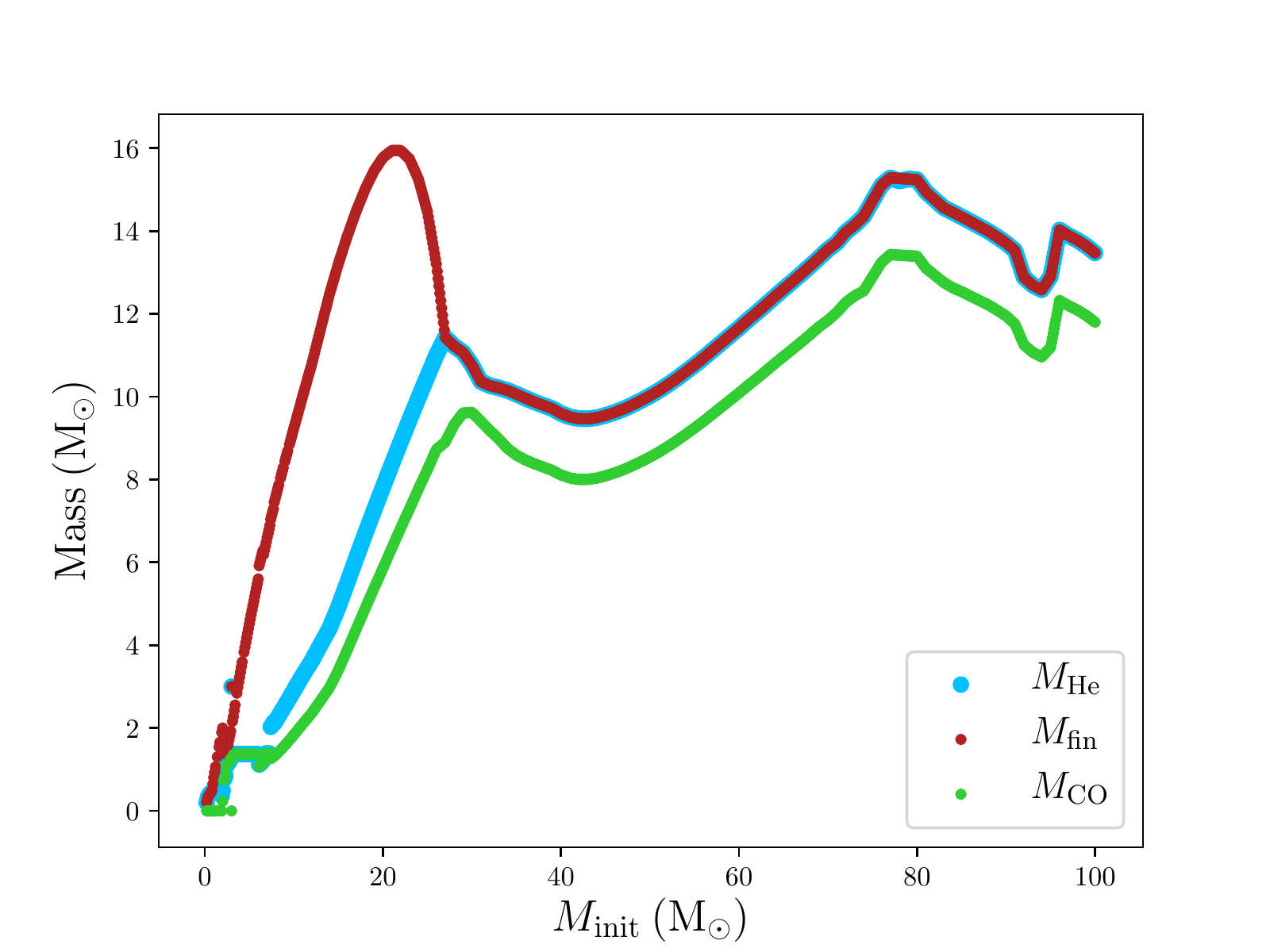}
    \includegraphics[scale=0.54,trim={0.4cm 0.1cm 1.6cm 0.5cm},clip]{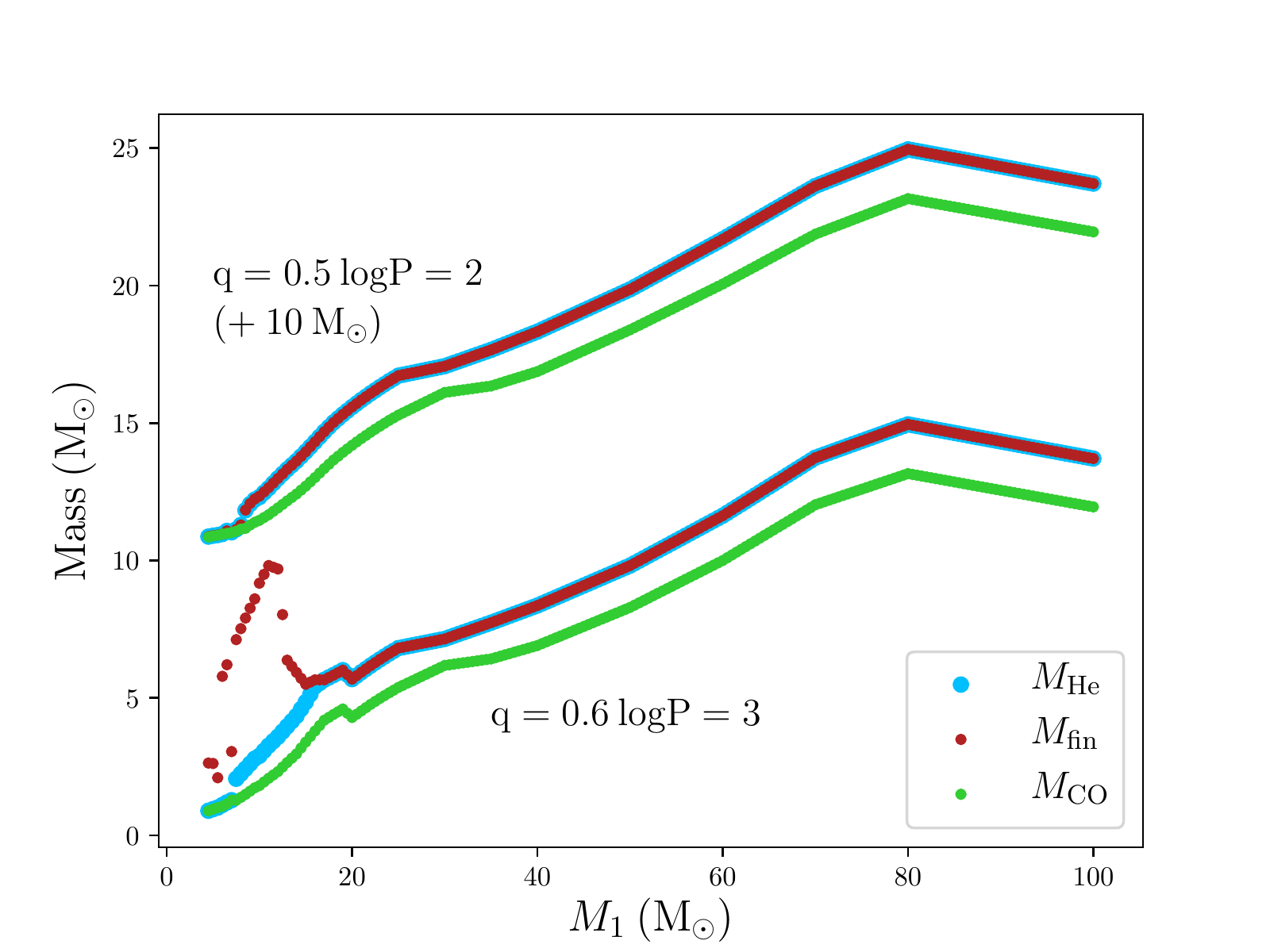}    
    \caption{A comparison of the helium core (blue), CO-core (green), and final mass (red) of stars for the single models (top) and a subset of the binary models (bottom). The q = 0.5, log P = 2 models have been offset by +10 \Msun~to have no overlap with the other models. The convergence of $M_{\rm fin}$ and $M_{\rm He}$ at $\sim$27 \Msun\ for single stars is caused by envelope stripping. Above this mass, all stars have lost their envelopes to winds and will ultimately experience collapse as bare cores. In binary models, the minimum mass at which a star loses its envelope depends heavily on the initial orbital period.}
    \lFig{mvm}
\end{figure}

\Fig{mvm} highlights key properties of these models, comparing \mfin, \mhe, and \mco~as a function of initial mass ($M_\mathrm{init}$ for single stars, $M_1$ for binaries) for the single stars in the top panel and sample subsets of binary stars, at fixed mass ratios and initial periods, in the bottom panel. All binary models experience some amount of mass loss via transfer onto a companion, leading to the final mass and He-core mass to either be much closer than in the single stars, or identical. After the turnover, the final mass converges with helium core mass and marks the minimum birth mass for complete envelope stripping. For single stars this happens at around 27 \Msun, and depending on the initial period it ranges between 4.5 \Msun\ ($\log P = 1.4$) and 30 \Msun\ ($\log P = 4$). After this convergence, \mfin, \mhe, and \mco~generally increase with initial mass, but non-monotonically due to the adopted wind prescriptions. BPASS uses the Wolf-Rayet mass loss prescriptions of \citet{Nug00}, which vary between WC and WN stars. The second mass peak around 80 \Msun\ is due to significant mass loss experienced by stars of higher mass on the main sequence. These stars bypass the red supergiant phase, instead evolving straight to Wolf-Rayet stars after the completion of core hydrogen burning. 

 Because BPASS evolves the full structure of stars in detail, we are able to keep track of \xc~in addition to \mco, both of which are measured at carbon ignition when evolution terminates. PS20 highlighted that the evolutionary history of a star writes itself into the core by changing the CO-core mass and its starting composition. It has also been shown that stars tend to fall in well-defined curves in \xc~- \mco~space set by the physics of that evolutionary history \citep[e.g., PS20, ][]{Sch21,Lap21}. By characterizing our models in this plane, we show for the first time how populations of stars have their core structures affected by binary interaction and wind mass loss. 
 
\Fig{xc} shows where the single and binary stars fall in this plane. There are two main sequences, one spanning between 1.4 - 14 \Msun\ in \mco~and 0.15 - 0.3 in \xc~that forms an upper bound, and the other ranging between 1.4 - 9.5 \Msun\ and \xc~of 0.10 - 0.25 bounding the points from the bottom. The lower curve, which we label the ``normal" sequence, is where stars fall if they retain some or all of their envelope. The upper curve, which we label the ``stripped" sequence, corresponds to stars which have lost all of their envelope, either by wind mass loss or mass transfer onto a binary companion. The characteristic curve shape is set by a few key aspects of stellar physics. In our set, the core mass generally increases with initial mass, and becomes less degenerate. The higher the core mass, the lower the central density, and the effect of C$^{12} (\alpha,\gamma)$ O$^{16}$ is stronger during core helium burning, leading to decreased \xc. For more degenerate lower mass cores, carbon burning via the 3$\alpha$ process is strong, increasing \xc. This is true whether or not the envelope has been stripped. However, the helium core ceases to grow when the envelope is completely stripped. Instead it recedes, leaving behind a carbon gradient and resulting in less helium being brought into the core. \xc\ increases, causing the offset between the two sequences. 

\begin{figure}
    \centering
    \includegraphics[scale=0.52,trim={0.1cm 0.0cm 1.6cm 1.3cm},clip]{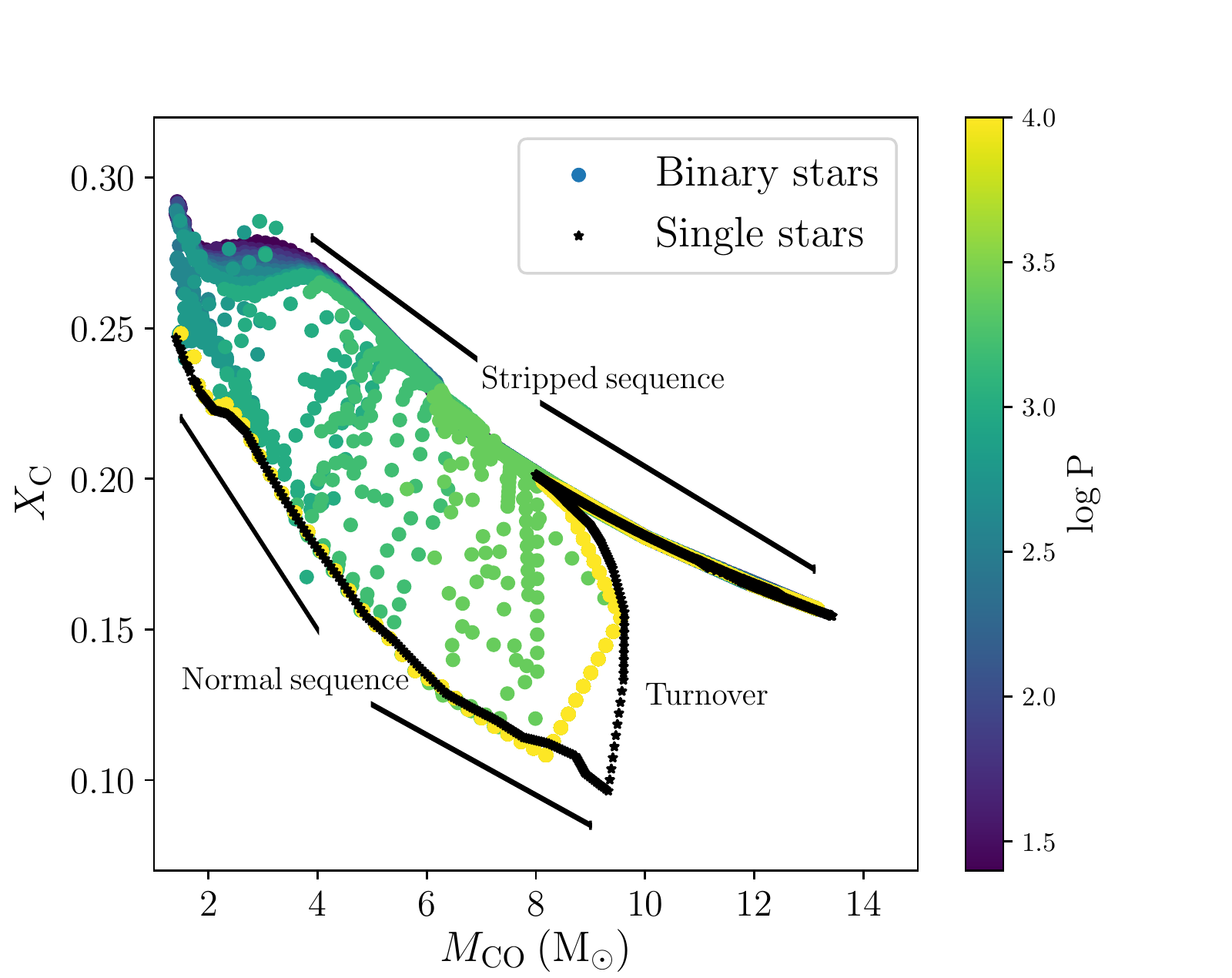}
    \includegraphics[scale=0.52,trim={0.0cm 0.0cm 1.6cm 1.4cm},clip]{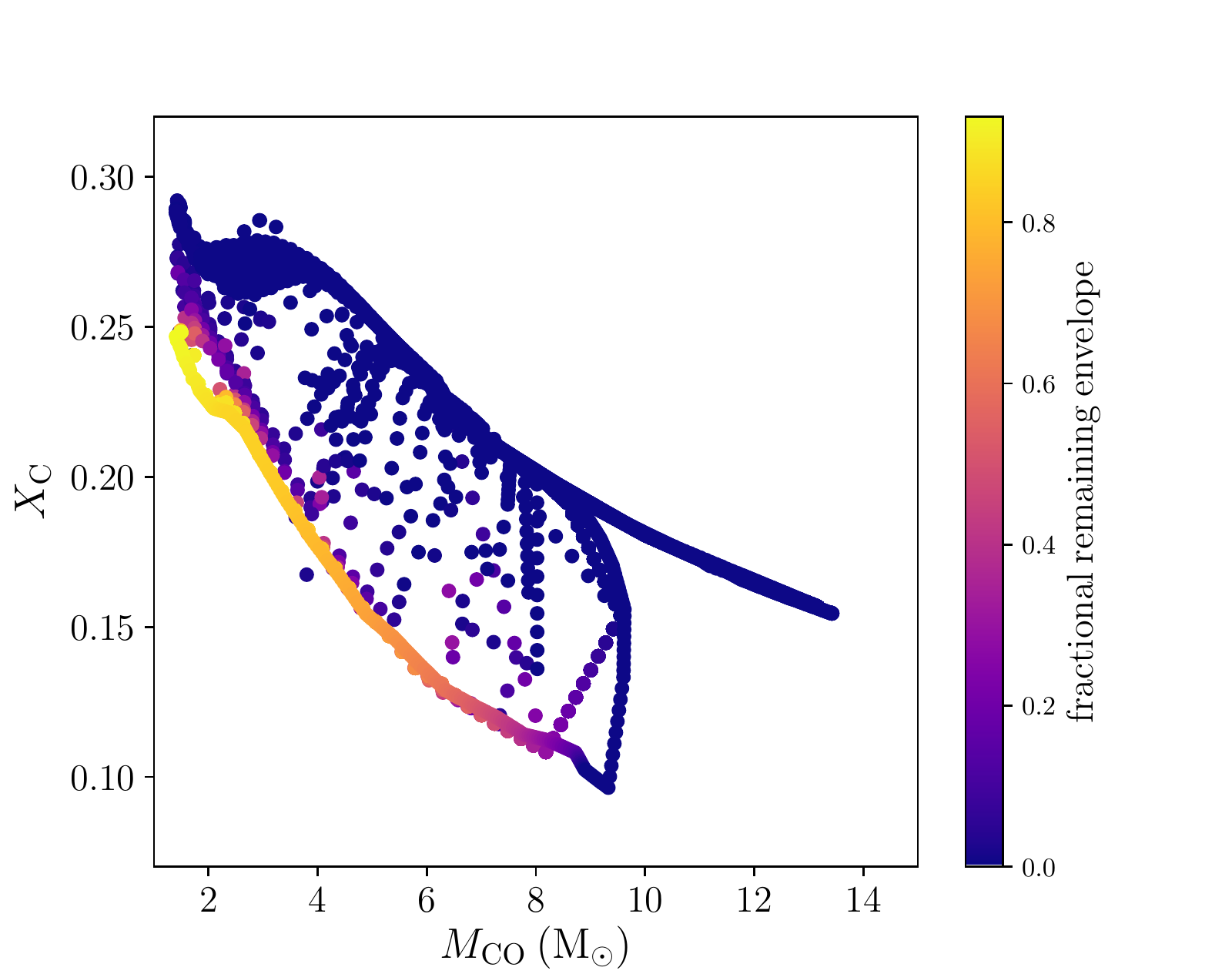}
    \caption{\xc~and \mco~values for the populations of single and binary stars from BPASS. Top: Binary models are color coded by their initial period, and all single stars are highlighted in black. Bottom: All single and binary models are color coded by the amount of hydrogen envelop mass retained at the time of carbon ignition. The upper bound (stripped sequence) of the models are made up of stars without any envelope, lost all through mass transfer onto a binary companion and/or wind mass loss. The lower bound (normal sequence) corresponds to the expected relationship between \mco~and \xc~for stars whose helium core evolves either in or effectively in isolation. The upward streams (turnover) connecting the two sequences, at around 10 \Msun\ for single stars and at lower values for binaries depending on the initial period, marks the transition from regular to stripped sequence.}
    \lFig{xc}
\end{figure}

Between the two sequences, we see turnovers in \xc. Each turnover roughly corresponds to a subset in our population, with turnover mass point corresponding to its lowest mass star which has its envelope entirely stripped. For the single stars, this is the 27 \Msun~model (\mco~$\approx 9$\Msun). With the binaries, envelope stripping can occur at much lower primary masses depending on the star's initial orbital period. The scattered binary points between the two sequences correspond to the turnover populations of binaries, e.g., stars with little or to envelope at the same $q$ and $\log P$, but varying $M_1$. A shorter initial period corresponds to a lower mass turnover. Each of these populations follow roughly the same trends as the single stars. Models of increasing initial mass will fall on the normal sequence until a model has its envelope entirely stripped. Then stars of higher initial mass will turnover onto the stripped sequence. 

Mergers can complicate the picture a little more. Though merged binary systems were removed from the sample population, an initial analysis revealed that they tended to fall in the exact same place as the single stars. We suspect this is due to when the stars merge. If stars merge on the main sequence, we expect them to evolve equivalently to single stars of their new mass, meaning they will lie on the normal sequence until wind stripping causes a turnover, where higher mass models fall on the stripped sequence. However, there was a separate group of models with values of \xc~which fell below the normal sequence, all of which were merger products. We suspect these are models which merged post-main sequence but cannot verify this since we only know the population's initial and final conditions. If these are post-main sequence mergers, it is worth investigating their impact on core structure. A preferential decrease in \xc~results in cores more difficult to explode and could impact the relative rates of supernovae and BH formation.   

\subsection{Predicting remnant masses and building the distributions}

 While BPASS is capable of evolving stars to neon ignition, properties of the stars were extracted at carbon ignition for a consistent application of various models considered in this study. We recorded the initial mass, final mass, helium core mass, CO-core mass, composition of the CO core, and the star's IMF weight, and applied them to estimate compact remnant masses through various prescriptions. We consider 1.4 \Msun\ to be the minimum $M_\mathrm{CO}$ to experience core--collapse \citep[e.g.,][]{Woo15}. For simplicity, we have completely ignored the effects of natal kicks, and our compact remnant distributions directly describe the final fate for each star in our population. Unless the NS mass is directly determined in gravitational form, we convert all baryonic NS masses into gravitational masses using the methods in \citet{Lat01} with an adopted NS radius of 12 km. BH masses remain baryonic, though the difference between baryonic and gravitational mass is small. All distributions are normalized to 1, and NS masses are grouped in bins of 0.1 \Msun\ and BH masses are grouped in bins of 0.5 \Msun.
 
To create the PS20 based distributions, we first took the \mco\ and \xc\ values for each star to obtain their presupernova structures. \mco\ was measured at the place where the helium mass fraction dropped to 10\%.  The CO cores of BPASS stars are not pure mixes of carbon and oxygen alone, like the idealized models from PS20 are, but the contribution to its initial composition from other elements is insignificant. The final fates for each star are determined by applying the two-parameter criterion of \citet{Ert16,Ert20} on the predicted final CO-core structures. This entails measuring $M_4$, the enclosed mass at the place where specific entropy equals 4, and $\mu_4$, a radial gradient in mass at the point of $M_4$, in the presupernova profile and comparing them to their W18-engine based results (using $k_1$ = 0.182 and $k_2$ = 0.0608). For stars that were deemed to produce a supernova we take the $M_4$ value as the baryonic NS mass, and for stars that failed to explode we adopt the helium core mass value as BH mass. Scenarios where the star explodes but leaves behind a BH were not explored. See section 3.2 of PS20 for further details on this sample implementation.
 
The PS20 models span 2.5 to 10 \Msun\ in $M_\mathrm{CO}$, a narrower range than the BPASS models. Since a successful neutrino-powered explosion above $M_{\rm CO} > 10$ \Msun\ is unlikely, we simply assume they all implode to produce a BH, and adopt $M_{\rm He}$ for its mass. For stars with $M_{\rm CO}$ between 1.4 and 2.5 \Msun, all stars are assumed to explode and produce a NS no matter the starting composition of the CO-core. For these stars the baryonic NS mass is determined by directly interpolating in $M_{\rm CO}$ over the results of the lightest models from \citet{Suk16}.
 
A second set of distributions were computed through the delayed, binary prescription laid out in section 4.3 of F12. This method estimates the remnant mass based on the CO-core mass and final mass of the star. The prescription first estimates a remnant's initial baryonic mass based on a star’s CO-core mass. Then additional mass is added to the remnant, accounting for fallback, based on both the CO-core mass and final mass of the star. The distinction between NS and BH is left to the user, as the prescription solely estimates mass of a compact remnant. We assign a cutoff in baryonic mass of 2.5 \Msun~($\approx$ 2.3 \Msun~in gravitational mass), with NSs having masses less than the cutoff and BHs having masses above. F12 do suggest baryonic-to-gravitational mass conversions through \citet{Lat89} for NS, and an arbitrary 10\% reduction for BHs, but we refrain from adopting these methods to remain consistent across prescriptions. We only convert NS masses through \citet{Lat01}, and keep the BHs masses in baryonic form. 

\Fig{Fry} shows how the predicted remnant masses compare between F12 and PS20 across the $M_{\rm CO}$ range of our BPASS models. The left hand panel compares the remnants produced by single stars and the right hand panel compares the remnants from binary systems. There is very little overlap between the two. The F12 NS masses increase with increasing \mco, ranging from 1.2 to 2.2 \Msun, whereas the PS20 NS masses mostly cluster between 1.2 and 1.6 \Msun. In the binary panel, we see an additional group of higher mass NSs predicted by PS20, created in an island of explodability at higher birth mass \citep[e.g.,][]{Suk14,Suk18}. In general, NS masses in F12 have a nearly unique correspondence with \mco, while in PS20 similar remnants can be made by stars that are far apart in birth mass, and conversely, stars of similar birth masses can result in different remnants as well.

The lowest mass BHs vary between prescriptions, with F12 predicting BHs down to 2.5 \Msun\ for both the binary and single models, where as the PS20 prescription predicts BHs down to 3.5 \Msun, the helium-core mass of the lightest imploding model. The highest mass BHs agree between both prescriptions. Additionally, there is a turnover in remnant mass around \mco\ = 9.5 \Msun, caused by envelope stripping, apparent in both prescriptions, but it is more pronounced in the F12 remnants. This turnover marks the transition between the two sequences shown in \Fig{xc}, with every remnant above the turnover coming from a stripped star. The BH masses from F12 also increase more quickly with \mco\ than in the PS20 models due to the dependence of the remnant mass on \mfin, which is more strongly affected by mass loss from winds or binary interaction than is \mhe. BH masses only agree for the highest mass stars because their envelopes have been completely stripped; \mfin\ = \mhe. In addition, more apparent in the binary panel, the PS20 models predict a population of NSs, due its consideration of core structure, for stars that were deemed to produce a BH according to F12.

\begin{figure}
    \centering
    \includegraphics[scale=0.54,trim={0.4cm 0.2cm 1.4cm 0.9cm},clip]{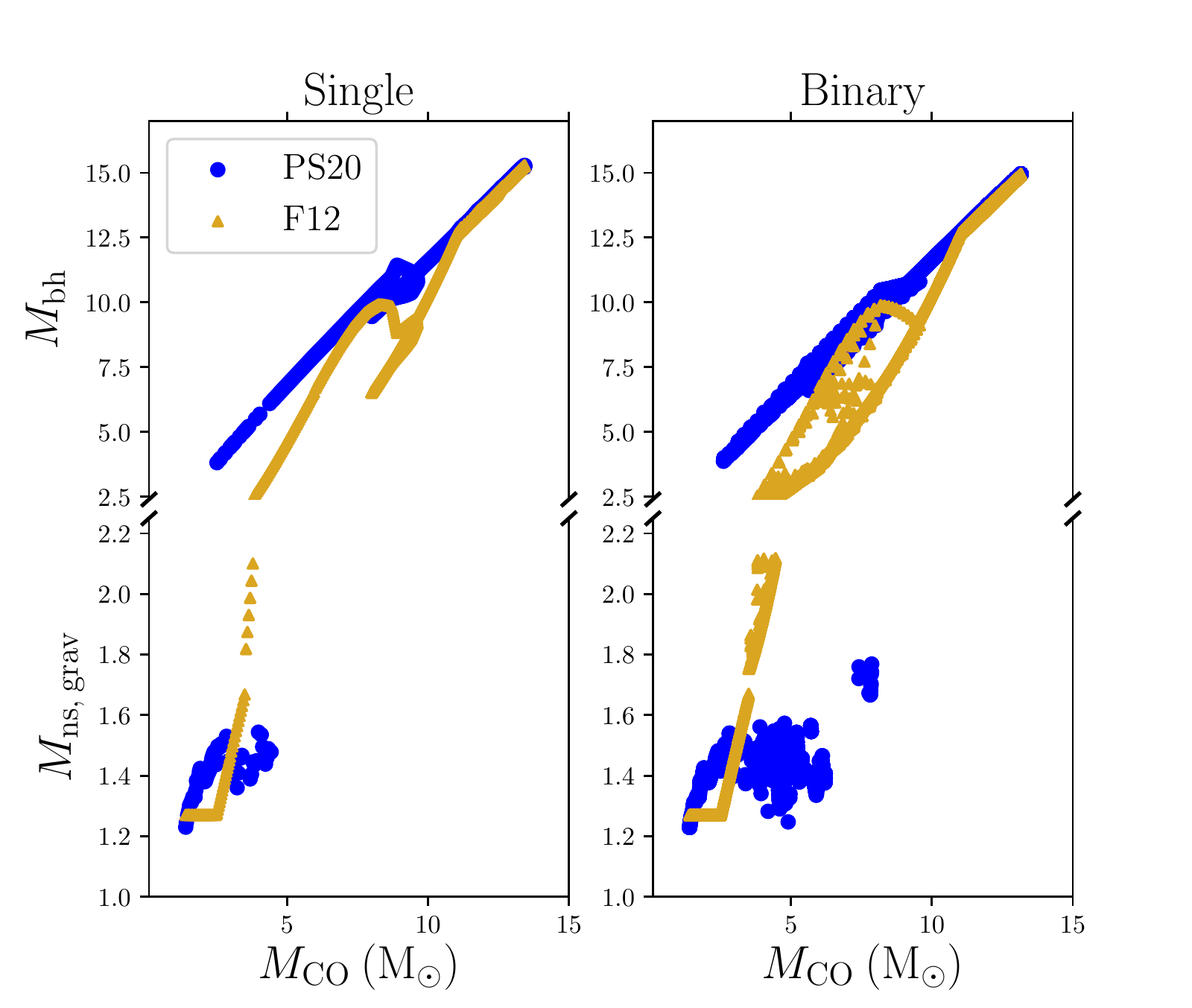}
    \caption{Remnant masses determined by the F12 (gold triangles) and PS20 (blue circles) prescriptions as a function of \mco~for each BPASS model, single (left) and binary (right), capable of reaching core--collapse. NS gravitational masses are shown on the bottom, and baryonic BH masses on the top; note the vertical axis break. The greater diversity in outcomes in the right panel demonstrates the impact of binary interaction on the core. The steady increase in remnant mass indicates that the F12 prescription effectively functions as a cut in \mco, separating the NSs (explosions) from the BHs (implosions).}
    \lFig{Fry}
\end{figure}

These discrepancies point to the key issue with the F12 method and any prescription that ignores late-stage evolution: the failure to consider the role presupernova core structure plays in determining final fate and remnant type \citep[e.g.,][]{Woo02,OCo11,Ugl12,Pej15,Suk16,Suk20}. That presupernova core structure, in turn, depends on how the CO-core evolves from carbon ignition to collapse, and the evolutionary path taken by the CO-core depends dominantly on its mass \textit{and} initial composition. But because BPS codes often terminate at carbon ignition, the stages of evolution most important in determining final core structure are precisely the stages of evolution disregarded in the models. Prescriptions which only predict final fate and remnant properties based on a star's mass disregard the critical impact on structure by the CO-core's starting composition. Two stars with identical CO-core masses but different core compositions could have dramatically different outcomes.

Finally, we apply the hybrid prescription. The benefit of this method is its compatibility with every BPS code, unlike the PS20 models, which only work with codes that employ active stellar evolution. We construct a similar table as in \citet{Rom20}, by taking the pre-supernova helium-core masses from \citet{Suk16} and their associated baryonic NS masses, which include any late time fallback. For models below 12 \Msun\ we adopt the Z9.6 explosion engine, and above we use the W18 engine. Non-exploding models were assigned \mhe~as their remnant mass. We then interpolated over the table using \mhe~from both the single and binary star BPASS models. If a helium core mass fell between two models with different outcomes, the remnant type of the lower bound model was adopted. If that remnant was a NS, then we adopt the NS mass from the constructed table. If that model resulted in an implosion, we adopt \mhe\ of the BPASS models as the remnant mass. Because the tabulated remnant masses already reflect outcome, explosion versus implosion, we do not need to impose a cutoff between the two.

\Fig{hyPS} compares the predicted remnant masses for both the PS20 and hybrid prescriptions. The slight discrepancy in NS masses from stars with \mco\ below 2.5 is due to interpolating over the remnant masses from \citep{Suk16} using \mco\ (for PS20) and \mhe\ (for hybrid). Because both methods adopt \mhe\ as the BH mass, the agreement between the two methods for BHs is very good. There are only slight discrepancies, where the hybrid method predicts a small island of explodability for the single star population at higher \mco, and the PS20 method does not. The location and extent of such islands depends on the models the user selects to build the table; see \citet{Suk18} for a discussion on the differences between underlying models used in PS20 and hybrid prescriptions. 

\begin{figure}
    \centering
    \includegraphics[scale=0.53,trim={0.2cm 0.3cm 1.2cm 1.0cm},clip]{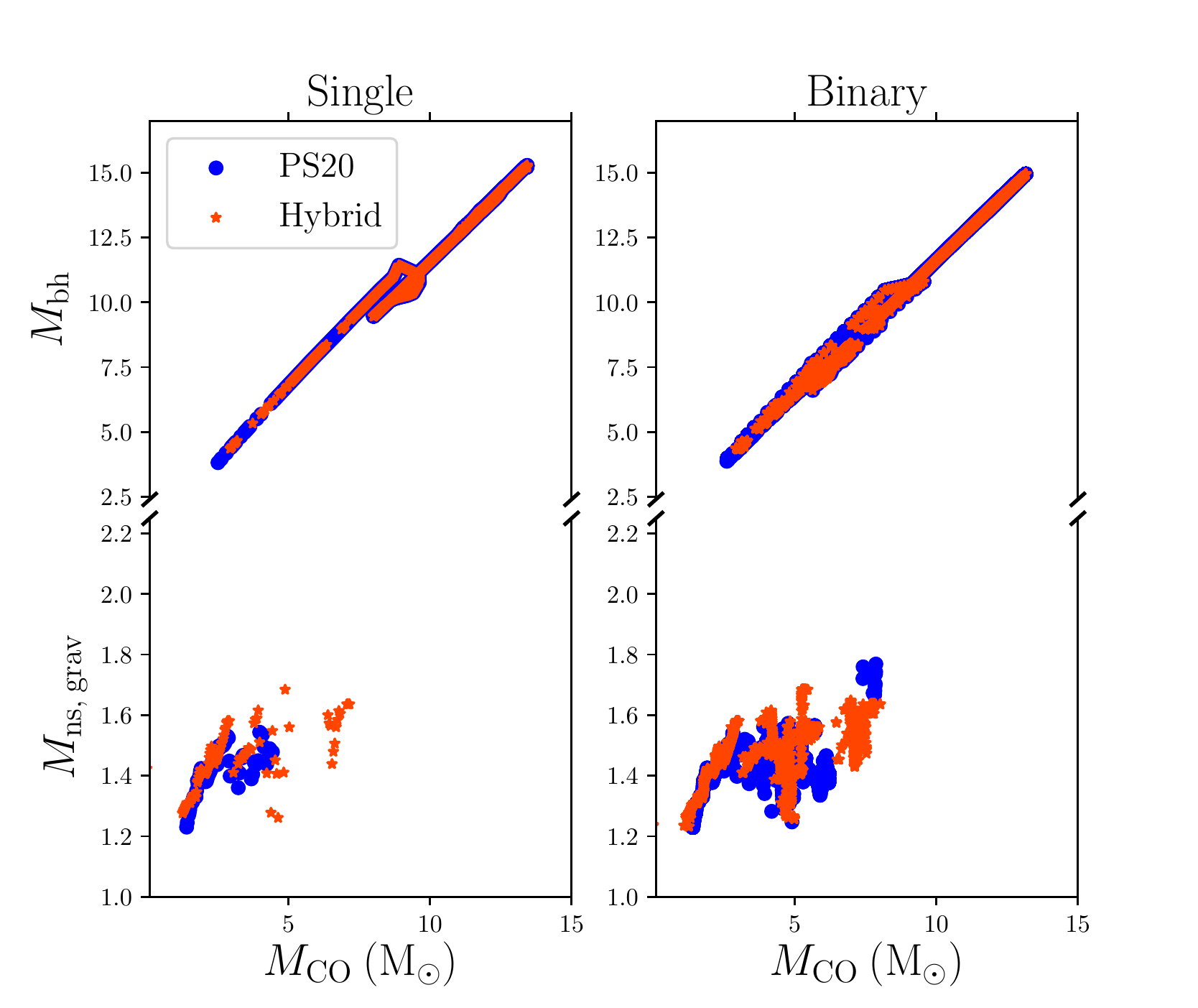}
    \caption{A comparison of the predicted remnant masses as a function of CO-core mass for the PS20 (blue circles) and hybrid (orange stars) prescriptions. The left panel shows the results from the single star models and the right panel, the binary models. The two prescriptions predict nearly identical BH masses but vary somewhat in NS mass predictions. The maximum \mco\ which produces a NS depends on the underlying models' islands of explodability, which depend on the adopted evolutionary physics. NS masses are gravitational, while BH masses are baryonic; note the axis break.}
    \lFig{hyPS}
\end{figure}

\begin{figure}
    \centering
    \includegraphics[scale=0.43,trim={0.0cm 0.0cm 0.8cm 1.0cm},clip]{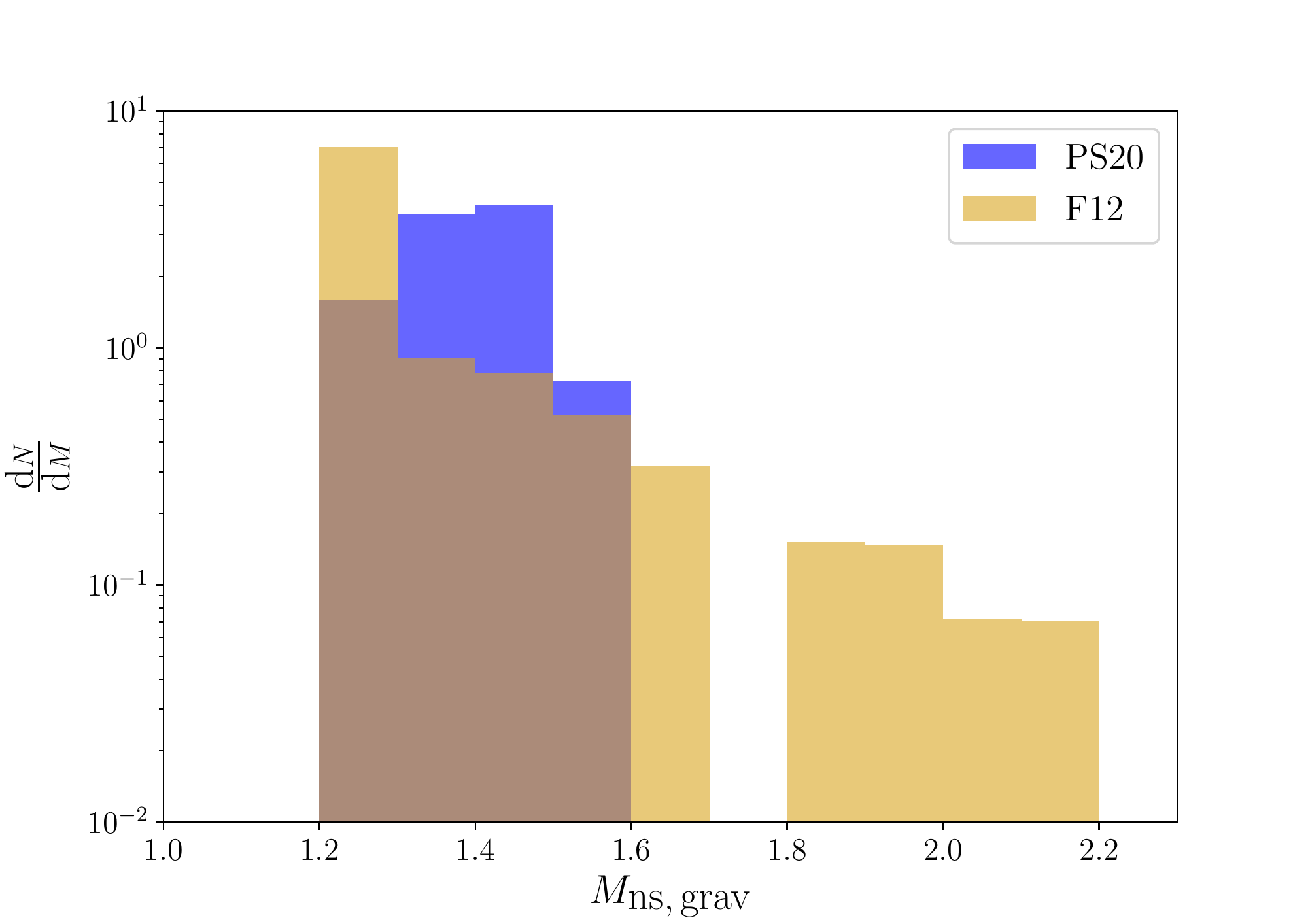}
    \includegraphics[scale=0.43,trim={0.1cm 0.0cm 0.8cm 1.0cm},clip]{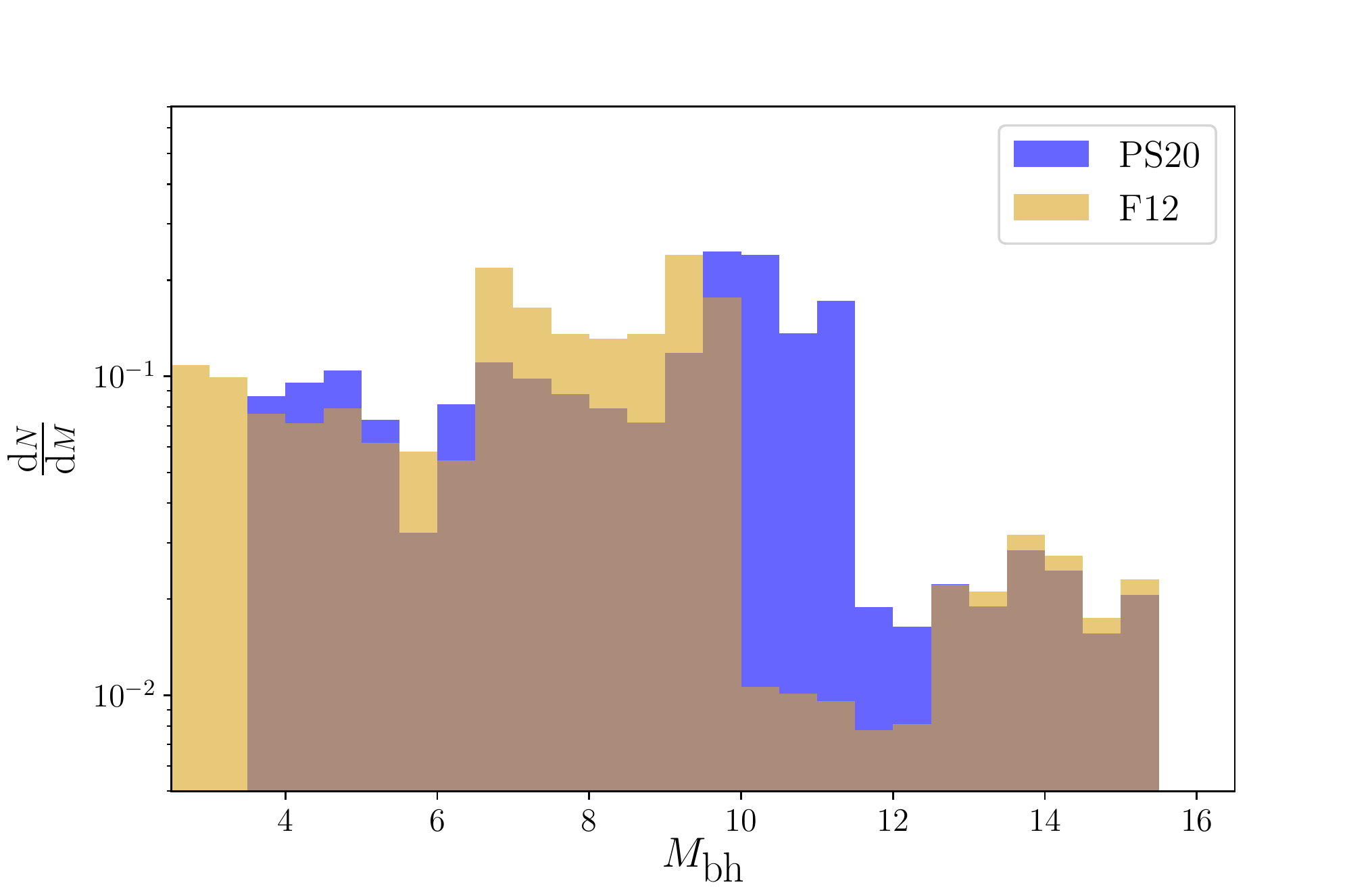}
    \includegraphics[scale=0.43,trim={0.0cm 0.0cm 0.8cm 1.2cm},clip]{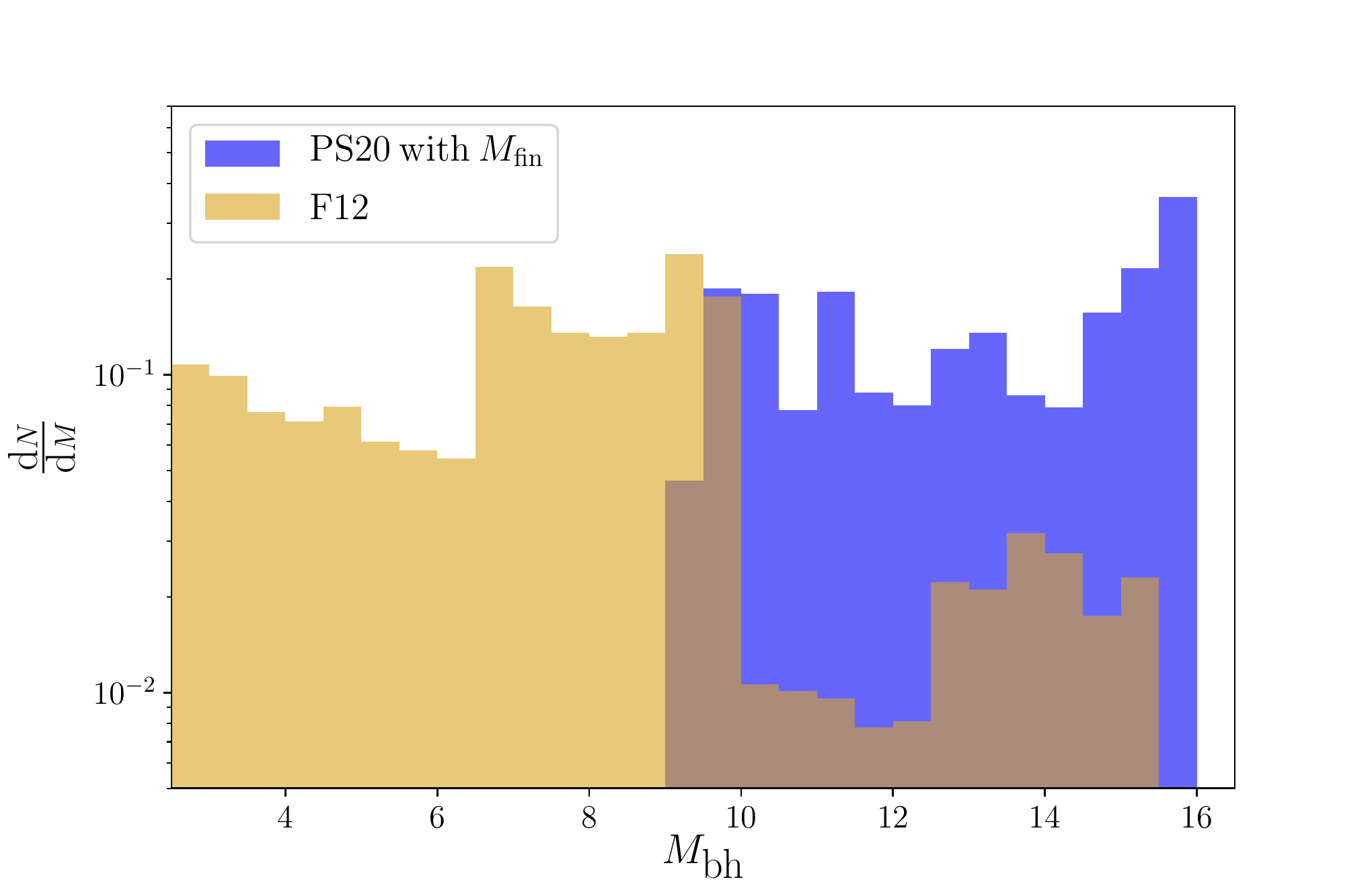}
    \caption{Compact object distributions for a population of 10$^6$ \Msun\ of single BPASS models determined by the PS20 (blue) and F12 (gold) prescriptions. The top panel shows the NS distribution. The center panel shows the black hole distribution. The bottom panel again shows the black hole distribution with the PS20 prescription adopting \mfin\ instead of \mhe~as the black hole mass. The intrinsically different shapes of the NS distributions point to the non-negligible differences in how each prescription determines its NS mass.}
    \lFig{sing}
\end{figure}

\section{Comparing the PS20 and F12 Distributions}
\lSect{res}
Differences between the two distributions are significant for both the single and binary populations. In this section we highlight the key properties between the two.

\subsection{Single star distributions}
\lSect{single}
Previous works employing combination of detailed stellar evolution calculations until iron core--collapse and calibrated neutrino-driven explosion models have shown that remnant distributions from single stars qualitatively match observed NS and BH mass functions \citep{Suk16, Suk18, Ert20}. However, these distributions have not been compared to those predicted from common prescriptions employed in BPS codes. We present below a brief comparison of the F12 and PS20 predicted compact remnant distributions based on single star population.

\subsubsection{Neutron star distribution}
\lSect{NS}
The top panel of \Fig{sing} shows the distribution of gravitational NS masses predicted by the PS20 and F12 prescriptions, revealing two key differences. The PS20 results show a narrow distribution ranging between 1.2 and 1.6 \Msun, and exhibit a sharp peak at 1.4 \Msun\ (IMF-weighted average mass of 1.39 \Msun), comparable to the observed peak of the NS mass function \citep[e.g.,][]{Oze12, Oze16} and to the predictions based on prior detailed calculations \citep[e.g.,][]{Tim96, Pej15}. In contrast, the F12 distribution has very different shape and predicts a substantially smaller value for the most common NS. The peak is located at 1.2 \Msun, the smallest value computed in the prescription, and sharply drops off with increasing NS mass, though the average NS mass is 1.35 \Msun. We expect this behavior because NS masses from the F12 prescription have a nearly monotonic dependence on \mco\ (\Fig{Fry}), which generally increases with initial mass (\Fig{mvm}) while ${dN}/{dM}$ decreases, resulting in more low-mass NSs. Since the F12 prescription only depends on \mco\ and \mfin\, it is most sensitive to the properties of the underlying stellar population. While the PS20 prescription is also sensitive to the underlying population, its dependence on core structure mitigates some of that sensitivity because stars with different initial masses can have similar core structure, and thus similar NS masses. There is also a gap in the F12 distribution caused by the transition between two different fallback mass calculations. The NS stars with masses above the gap have their fallback masses calculated depending on both \mfin\ and \mco, whereas the fallback mass for NSs below the gap depends on CO-core mass alone. 

\subsubsection{Black hole distribution}
\lSect{BH}
There is more qualitative agreement between the F12 and PS20 prescriptions in the BH distributions than in the NS distributions. In accord with the predictions from \citet[][]{Suk16}, both distributions shown in the middle panel of \Fig{sing} peak around $\sim$10 \Msun, though the IMF weighted average remnant masses are 8.59 and 7.52 \Msun\ for the PS20 and F12 distributions respectively. Both distributions also nearly match in predicted frequencies of BHs above 12.5 \Msun. The progenitors of these systems have their envelopes entirely stripped by winds, causing the final stellar mass and \mhe\ to be equivalent (stars above 27 \Msun, see top panel of \Fig{mvm}). 

But below 12.5 \Msun\ the two distributions are most discrepant because of the differences in how BH mass is determined. The PS20 distribution is only sensitive to the underlying population's He-cores since \mhe\ is adopted as $M_\mathrm{bh}$. It experiences a drop in frequency starting at 11.5 \Msun, the local maximum mass at which the He-cores of the single star population reach before turning over. The He-cores do not grow again until they shrink to 9 \Msun, coming from stars with initial masses above 43 \Msun. But these stars are so massive, the IMF weighting strongly disfavors them, meaning an overall decrease BH frequency for BHs above 11.5 \Msun.

The F12 models are also sensitive to these effects because, although the F12 distribution determines BH mass from \mfin, it is equivalent to helium core mass (\mfin\ = \mhe) for stars with birth masses above 27 \Msun. However, \mfin\ turns over before \mhe\ does. At 21 \Msun\, the single stars reach a global maximum in \mfin, then drop off (\Fig{mvm}). Decreasing \mfin\ decreases $M_\mathrm{bh}$, leaving a deficit of BH masses down to 10 \Msun\, instead of 11.5 \Msun\ in the PS20 distribution.

Lower-mass BHs also exhibit an important difference between the two distributions, namely the prediction of lower-mass mass gap remnants. The simple implementation of PS20 in this study ignores all fallback (remnant mass is either $M_4$ or \mhe\ depending on the outcome), the smallest BH mass is effectively the smallest Mhe that implodes (3.5 \Msun). However, in F12 the remnant mass continuously increases with Mco, and the distributions predict a substantial population of BHs as low as 2.5 \Msun, the adopted lower limit. Though there is some evidence for mass gap BHs, such as those predicted by \citet{Ert20} and the observed BH - red giant binaries from \citet{Tho19, Jay21}, as well as new observational techniques to detect them \citep[][]{Jay21}, it remains unclear how common they are and what are their origins.

In addition to comparing the distributions from the two prescriptions as is, we also augment a version of PS20 prescription to adopt the final stellar mass instead of \mhe\ as the BH mass, shown in the bottom panel of \Fig{sing}. We do this first and foremost because of the uncertainty in how much of the total stellar mass ultimately ends up in the BH. For the loosely bound cool envelopes of red supergiants, \citet{Lov13} suggest that neutrino emission could unbind the envelope, thus $M_{\rm BH} \sim \mhe$. However, for the more tightly bound compact envelopes of blue supergiants, the neutrino emission might not be sufficient to eject any part of the envelope. In this instance, \mfin\ would be a better estimate of the BH mass. Ideally, for a more careful treatment one would need to use the envelope structure at core--collapse (or carbon ignition, since the envelope does not have time adapt to the changing core) and consider the binding energy profile in order to adopt \mhe\ or \mfin\ or somewhere in between for the BH mass. 

In the absence of this knowledge, we show here how the BH distribution changes with \mfin\ adopted as the BH mass. This also provides a slightly better comparison to the F12 prescription, which uses the final mass of the star to determine the amount of fallback mass on higher mass CO-cores, where the most massive cores (>11 \Msun) produce BH masses equivalent to the final mass of the star and stars with 6 < \mco\ < 11 \Msun\ produce BHs with masses set by some \mfin-dependent fraction of the final mass. Stars with \mco\ < 6 \Msun\ have either a fixed or \mco-dependent amount of fallback. This leads to a greater discrepancy between the PS20 and F12 distributions because F12 produces a spectrum of fallback masses whereas PS20 either adopts all or none of the envelope mass in addition to \mhe. For the PS20 models, every star which retained some or most of its envelope shifts upward in mass by about 5 \Msun, which dramatically increases the frequency of high mass BHs. This new distribution is comparable to the equivalent $M_{\rm BH}=M_{\rm fin}$ predictions of \citet{Suk16}.

\subsection{Binary star distributions}
\lSect{Bin}
Because binary interaction can substantially change the properties of the CO-core, we expect to see deviations from the single star distributions, such as higher mass remnants. Below we examine the extent to which binary interaction shapes the distributions.

\subsubsection{Neutron star distribution}
\lSect{NSB}
\begin{figure}
    \centering
    \includegraphics[scale=0.43,trim={0.0cm 0.0cm 1.1cm 1.0cm},clip]{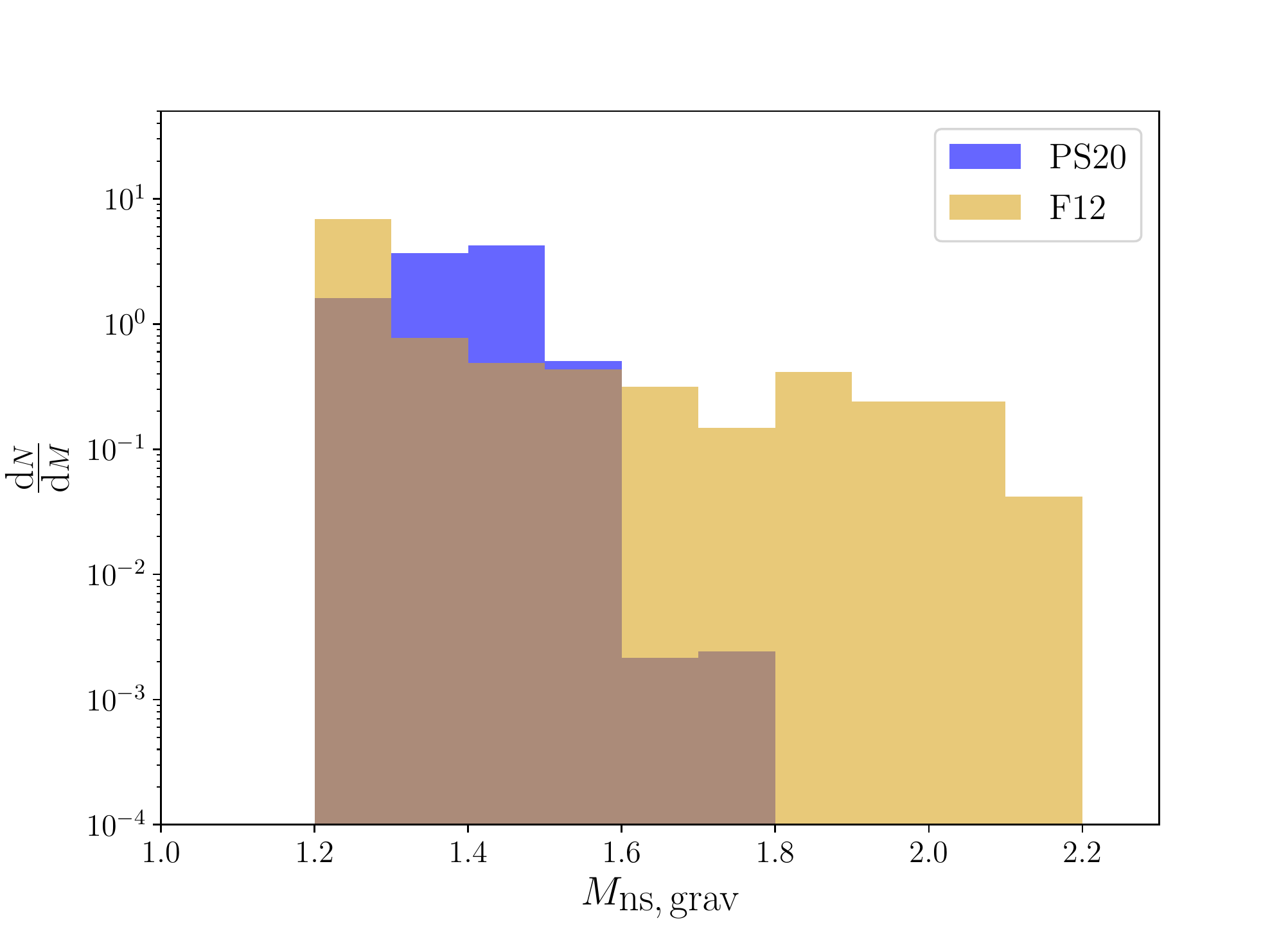}
    \caption{NS distribution from stars originally in binary systems determined by the F12 (gold) and PS20 (blue) prescriptions. Because the F12 prescription predicts increasing remnant mass with \mco~and initial mass, the distribution extends to higher NS mass.}
    \lFig{NSbindist}
\end{figure}

The differences between the F12 and PS20 prescriptions are most apparent in the NS distribution from binary stars. Again, the PS20 distribution peaks near a NS mass of 1.4 \Msun\ and retains a more normal shape. The F12 distribution is more extended, with a slight peak at 1.2 \Msun. Still, both prescriptions produce average NS masses of 1.38 \Msun.

As shown in \Fig{NSbindist}, the PS20 distribution now extends to 1.8 \Msun. The higher mass NSs come exclusively from stars with \mco\ $>$ 7.5 \Msun. These stars happen to fall in an island of explodability in the PS20 models where the $M_4$ values are higher than usual. But just having a higher \mco\ does not necessarily entail a higher NS mass. Even NSs which come from progenitors with \mco\ greater than those of the single star population (\Fig{Fry}) still produce NSs with masses between 1.2 and 1.6 \Msun. This is because of how the star responds when its envelope is entirely stripped, either by extreme wind mass loss or mass transfer onto a companion. The lack of new helium being brought into the core combined with the carbon gradient left behind by the receding convective core results in higher \xc~ for a given \mco. Having a higher \xc\ extends the range of \mco\ which can produce NSs.

The F12 prescription is not sensitive to these changes in \xc, instead reflecting binary interaction only in changes to \mco\ and \mfin. The binary NS distribution still extends to 2.2 \Msun\ and, up to 1.8 \Msun\, it is nearly identical in frequency to the single F12 NS distribution. However, mass loss from binary interaction has shrunk core masses for some stars, leading to an over production of higher mass NSs (1.8-2.2 \Msun), hence the higher frequency in this mass range. Combinations of \mco\ and \mfin\ capable of producing a NS of a given mass can now come from a broader range of initial masses of their progenitors depending on how much mass a star loses or gains from binary interaction. The end result is a near continuous distribution of NSs.

\subsubsection{Black hole distribution}
\lSect{BHB}
The binary distributions are similar to the single star ones, just with slight reduction in maximum BH mass, 15\Msun~instead of 15.5 \Msun, caused by extreme mass loss. However, the highest mass BHs from binaries slightly increase in frequency over those from single star progenitors. This is caused by shallower decrease in \mfin\ and \mhe\ in the binary models for stars above 80 \Msun\ (see \Fig{mvm}). The PS20 models show no strong peak whereas the F12 distribution tends to favor lower mass BHs, again predicting high frequency of remnants down to 2.5 \Msun. The average remnant masses are similar to the single star distribution, 8.16 \Msun\ for the PS20 distribution and 7.33 \Msun\ for the F12.

Both the PS20 and F12 distributions experience a stark drop in frequency for BHs above 10 - 11 \Msun\ depending on the distribution, again caused by the turnover in \mhe, still present in the binary models, weight penalties from the IMF, and, for the F12 prescription, the turnover in \mfin\ for some of the binary populations (see details in \Sect{BH}).

When the PS20 prescription is changed to adopt \mfin~as the BH, the remnant masses shift higher (bottom panel of \Fig{BHbindist}). Because the stars which experience envelope stripping span a much wider range for binary systems than single stars, the PS20 distribution does not shift as dramatically as in the bottom panel of \Fig{sing}. However, the remnant mass range above 11 \Msun\ increases in frequency as the extra mass from the envelope is added back into total BH mass, leading to higher mass BHs coming from lower initial-mass progenitors in addition to the stars already populating this region (top panel of \Fig{BHbindist}). Because PS20 adopts \mfin\ as the BH mass for all BHs, the distribution extends to a higher maximum BH mass, which comes from a star with a lower initial mass that retains much of its envelope. In the F12 distribution, the highest mass cores (\mco\ > 11\Msun) are the ones which adopt \mfin\ as the BH mass, but these stars are also the most stripped; in most cases \mfin\ = \mhe, leading to lower mass BHs. 

\begin{figure}
    \centering
    \includegraphics[scale=0.42,trim={0.0cm 0.1cm 0.7cm 1.0cm},clip]{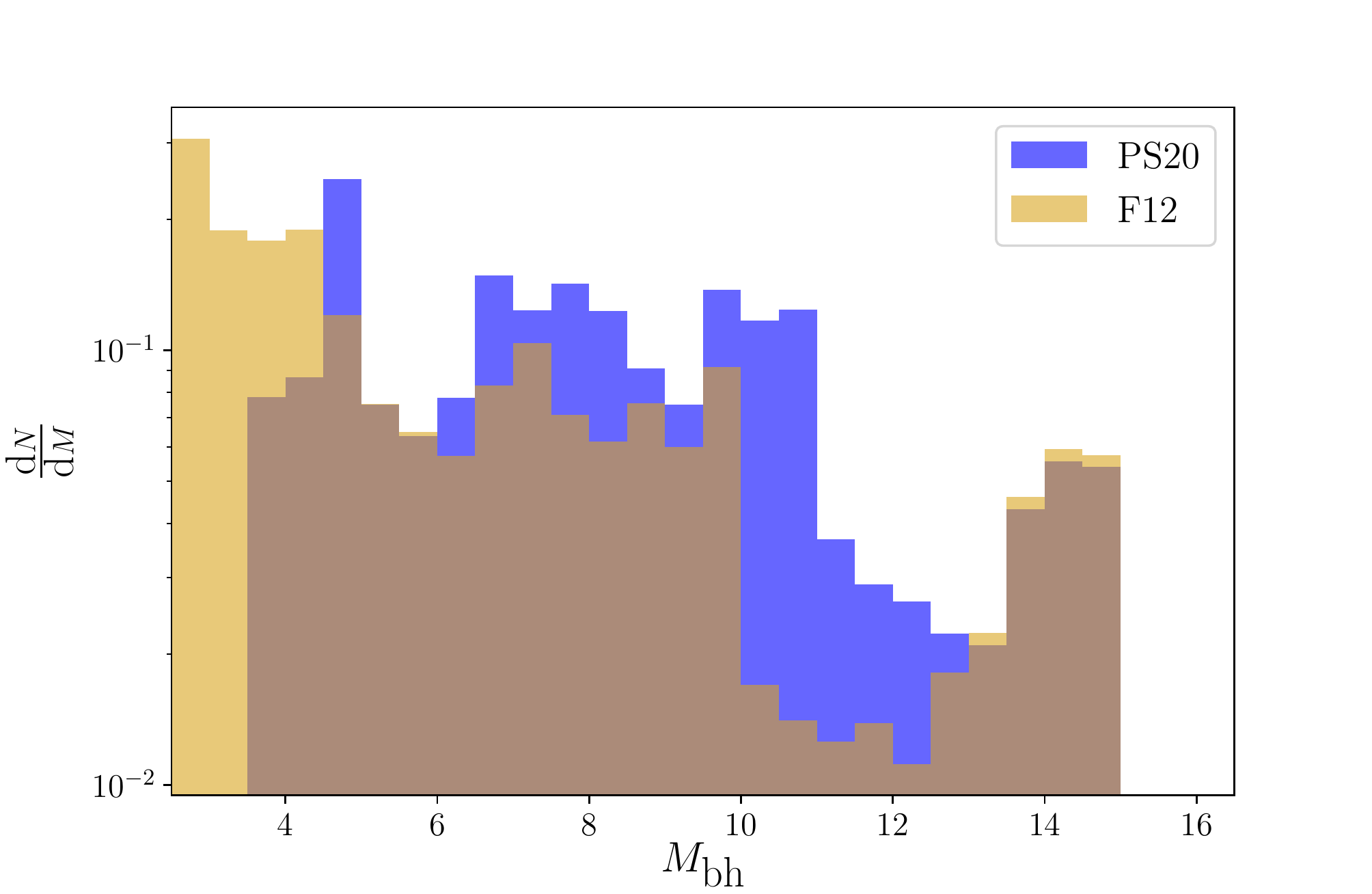}
    \includegraphics[scale=0.42,trim={0.0cm 0.1cm 0.8cm 1.0cm},clip]{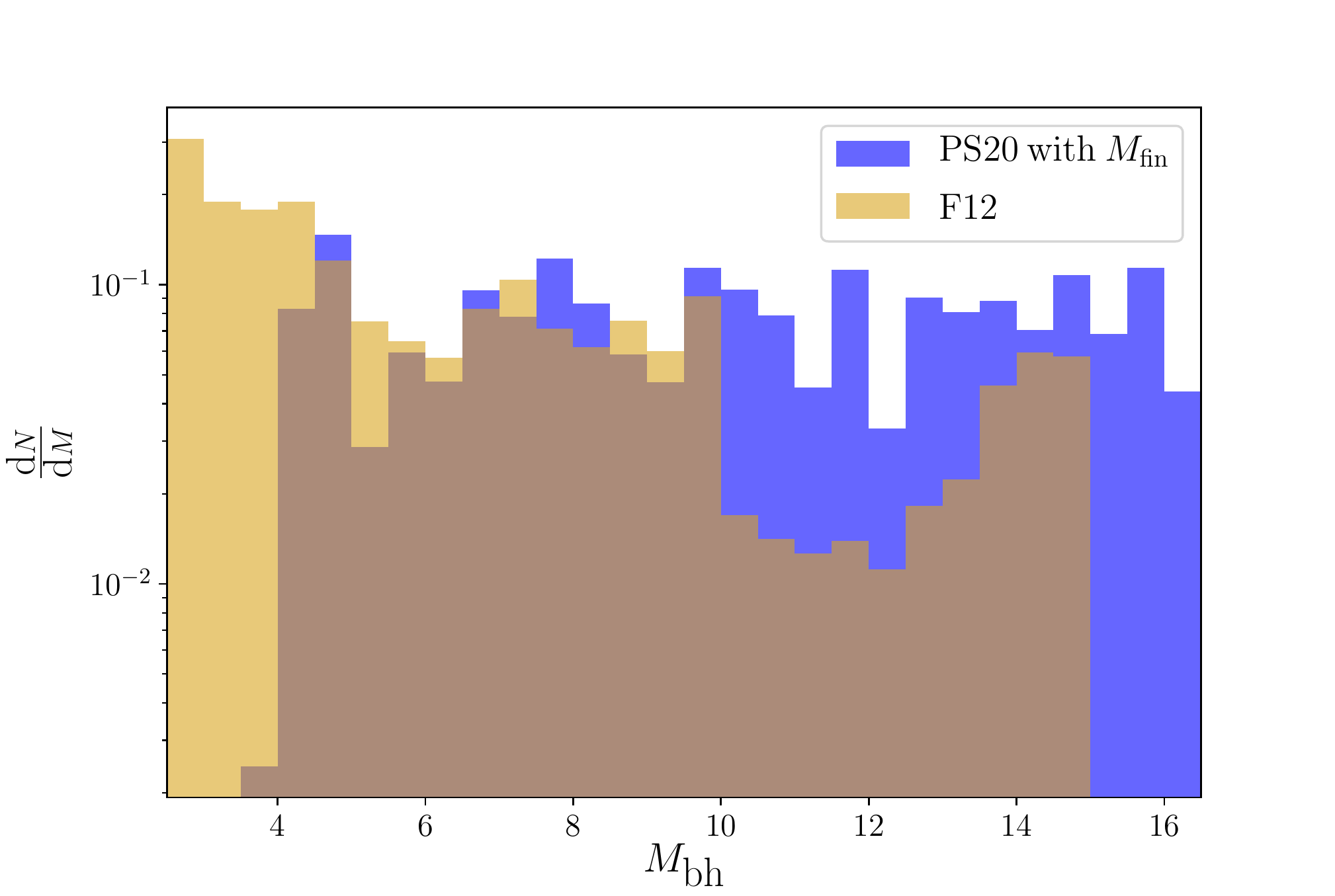}
    \caption{BH distributions determined by the F12 (gold) and PS20 (blue) prescriptions from the binary models. The top panel shows the two prescriptions as is whereas the bottom panel shows the two distributions if PS20 adopted the final stellar mass as the BH mass. Adoption of \mfin\ by PS20 fills out the drop in frequency for BHs above 11 \Msun and increases the maximum mass BH.}
    \lFig{BHbindist}
\end{figure}

\section{Comparing the PS20 and hybrid distributions}
\lSect{hybrid}

Because both methods predict remnant masses based on a star's presupernova core structure and are built from the results of the same explosion models, we expect the PS20 distributions to agree much more closely to the hybrid approach than F12. Below we discuss the key differences between the single and binary NS and BH distributions. 

\subsection{Single star distributions}
\lSect{hysin_res}

\Fig{hysin} shows both the NS and BH distributions predicted by the PS20 and hybrid prescriptions. The two methods largely agree both in predicted frequency and remnant mass range. The IMF weighted average remnant mass from each is comparable, 1.39 \Msun\ for the PS20 distribution and 1.4 \Msun\ for the hybrid. The \citet{Suk16} models do include fallback, but typically proto-NSs only gained a few hundredths of solar mass. 

There is also little disagreement between the two methods in the predicted BH distribution, and the PS20 and hybrid prescriptions predict comparable average remnant masses, 8.67 and 9.22 \Msun\ respectively. All BHs above 9\Msun have nearly identical predicted frequencies, and below, discrepancies are minor. The drop off due to the island of explodability are located at different masses, between 8.5 and 9 \Msun\ for hybrid, and between 5.5 - 6 \Msun\ for PS20 due to the differences in the underlying stellar models. We do not consider these dips absolute in their location, they are sensitive to both the adopted stellar and supernova physics (see section 3.2 in PS20). Finally, the lower limit on BH mass varies slightly between the two, 3.5 \Msun\ for the PS20 distribution and 4 \Msun\ for the hybrid due to slight differences in the explosion landscape. Here we note that had we used the \citet{Ert20} results to built the hybrid method, instead those of \citet{Suk16}, the BH distribution would have a substantial population of mass gap remnants.

\subsection{Binary star distributions}
\lSect{hybin_res}

The binary distributions for both the NSs and BHs experience the similar discrepancies as in the single models. The hybrid NSs extend up to 1.7 \Msun\, a slightly lower mass than the PS20 models, but still show similar behavior: clustering around a peak between 1.4 - 1.5 \Msun. The IMF averaged NS masses remain largely the same as well, 1.38 and 1.39 \Msun\ for the PS20 and hybrid distributions respectively.

The BH distributions are largely the same too. Again, they agree above 9 \Msun. Like in the PS20 and F12 distributions, there is a slight increase in frequency for the highest mass BHs over the ones from the single star progenitors. This is again caused by the shallower decrease in \mfin\ and \mhe\ in the binary models for stars above 80 \Msun\ (see \Fig{mvm}). The dip at 8.5 \Msun\ is caused by the island of explodability mentioned in \Sect{hysin_res}. And below 8.5 \Msun\ the distributions are comparable in their predicted frequencies but do not agree completely. Still, the average remnant mass remains comparable, 8.16 \Msun\ for the PS20 distribution and 8.43 \Msun\ for the hybrid.

\begin{figure}
    \centering
    \includegraphics[scale=0.42,trim={0.0cm 0.0cm 1.1cm 1.0cm},clip]{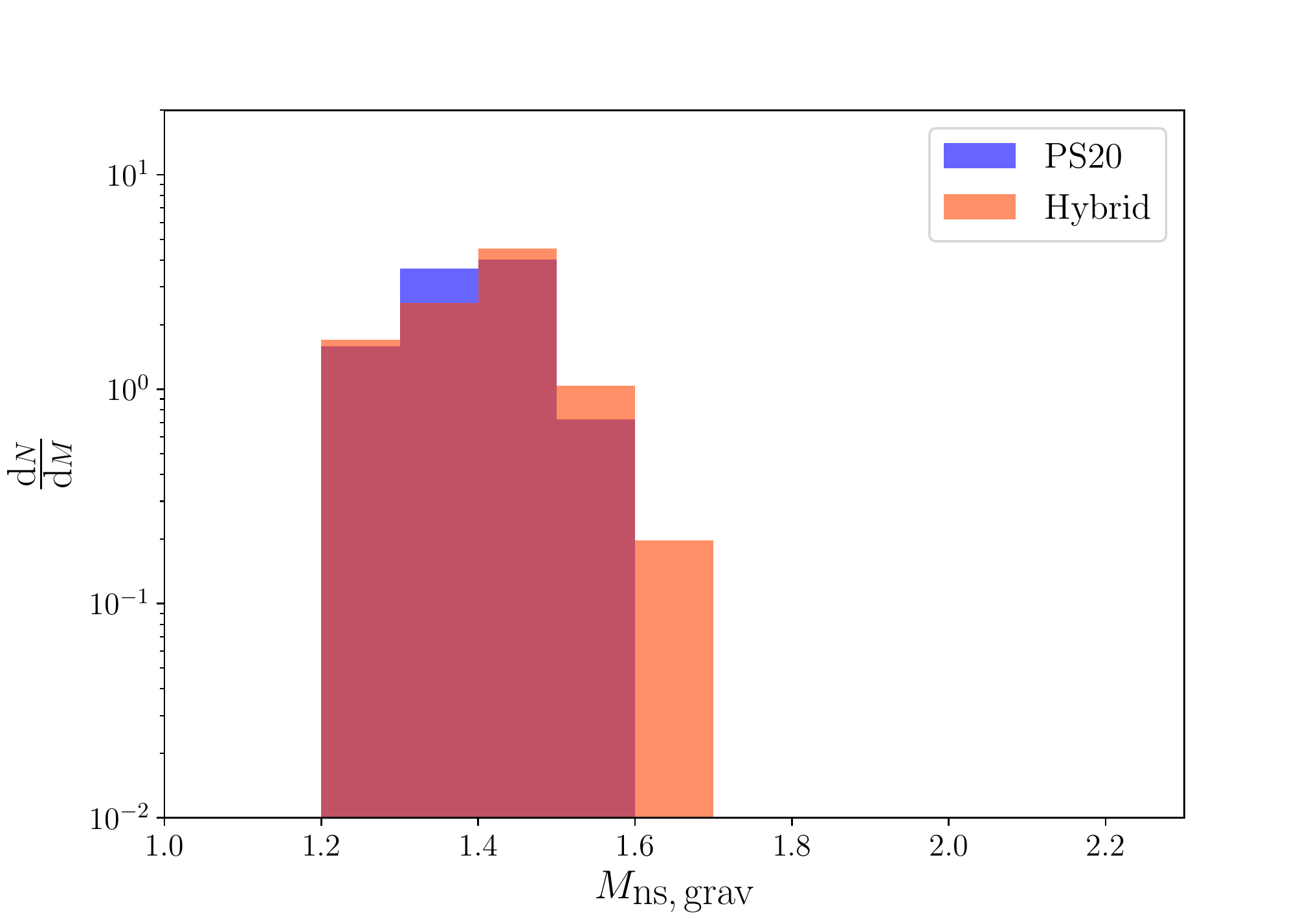}
    \includegraphics[scale=0.43,trim={0.0cm 0.0cm 1.1cm 1.0cm},clip]{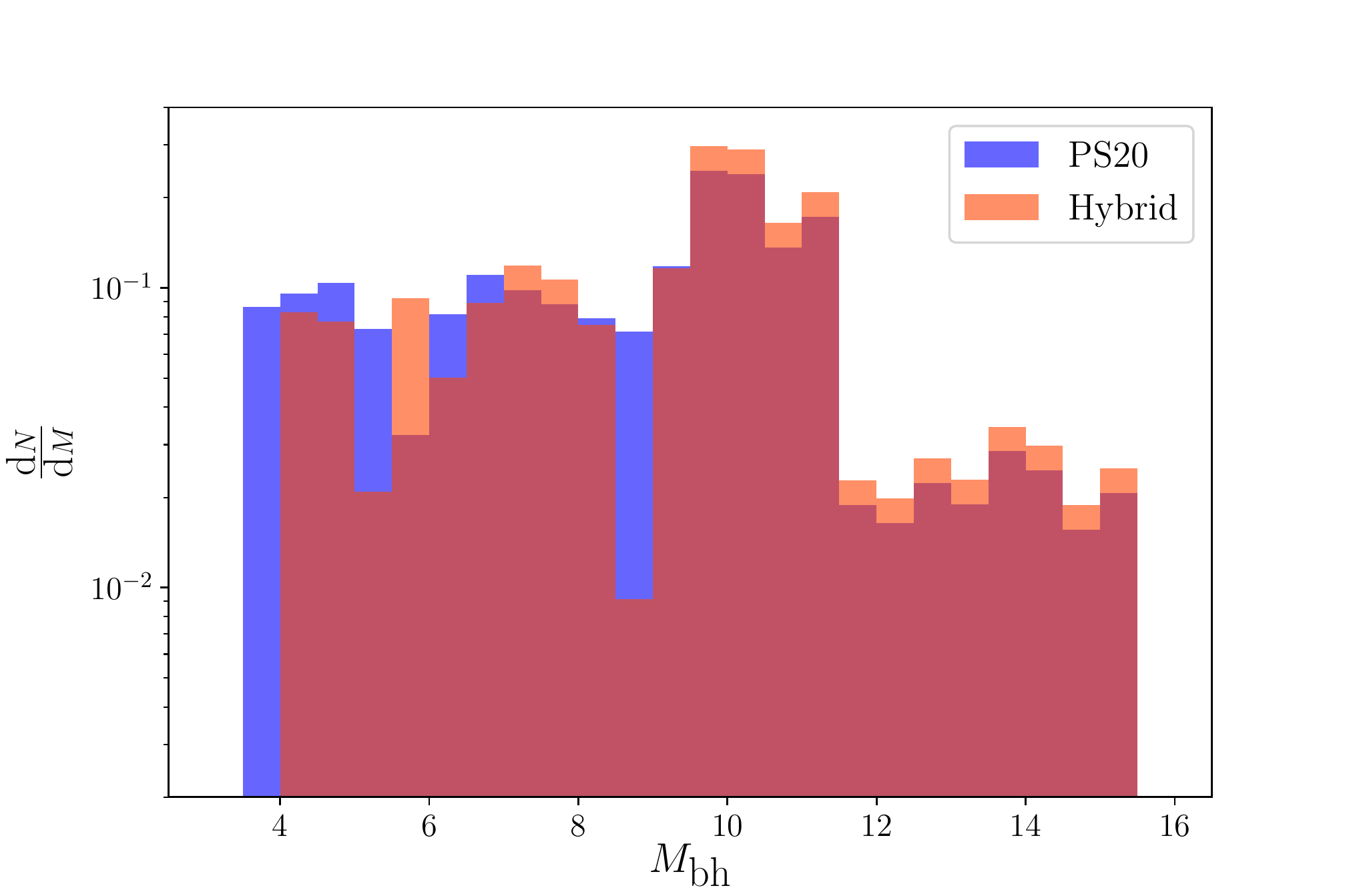}
    \caption{The predicted distributions from single progenitors using the prescriptions from PS20 (blue) and the hybrid approach (orange) for NSs (top) and BHs (bottom). Because both methods predict remnant masses based on the stars' presupernova core structures, the overall agreement between the two distributions in shape, frequency, and mass range is very good.}
    \lFig{hysin}
\end{figure}

\begin{figure}
    \centering
    \includegraphics[scale=0.43,trim={0.0cm 0.0cm 1.1cm 1.0cm},clip]{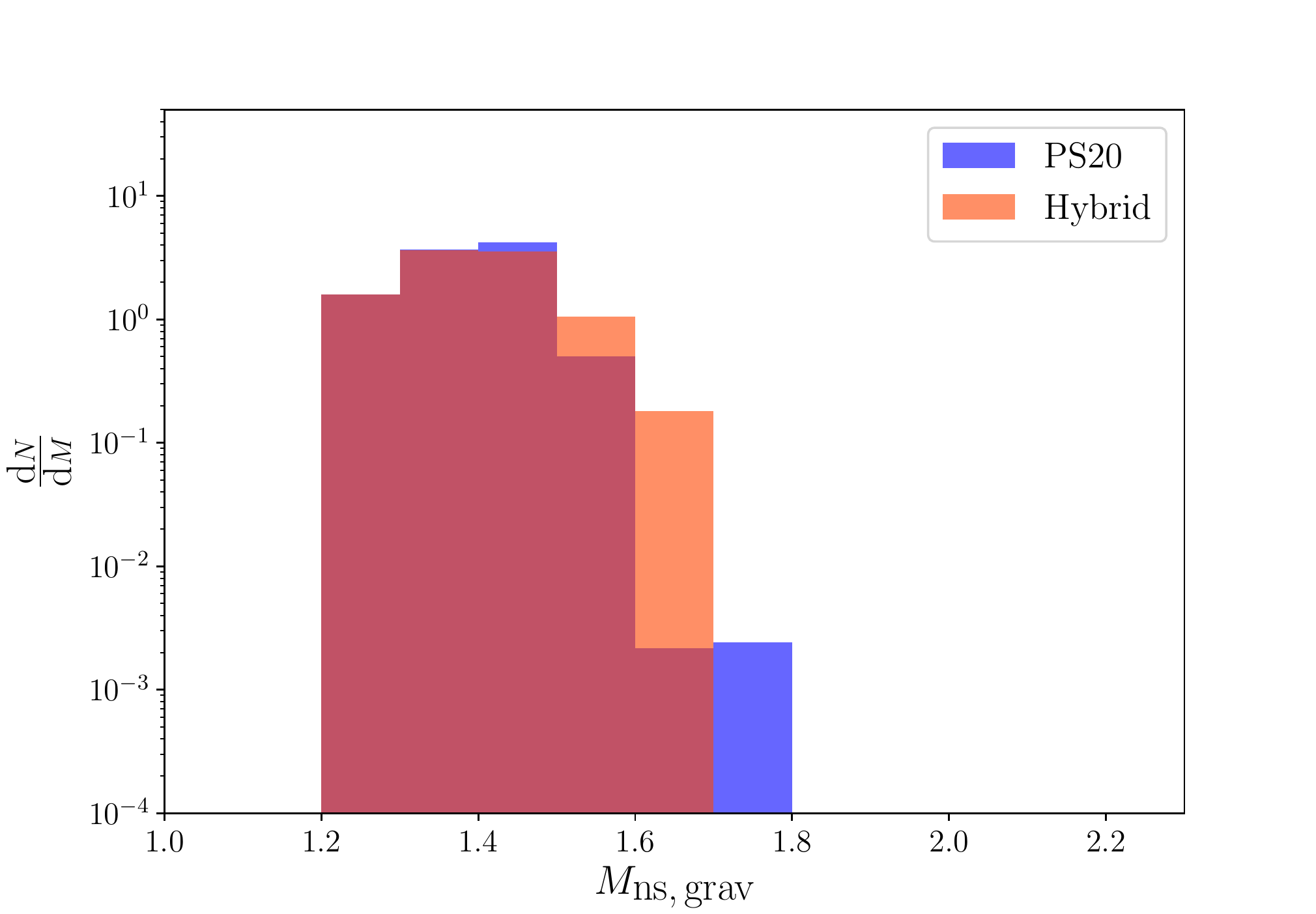}
    \includegraphics[scale=0.44,trim={0.1cm 0.0cm 1.1cm 0.9cm},clip]{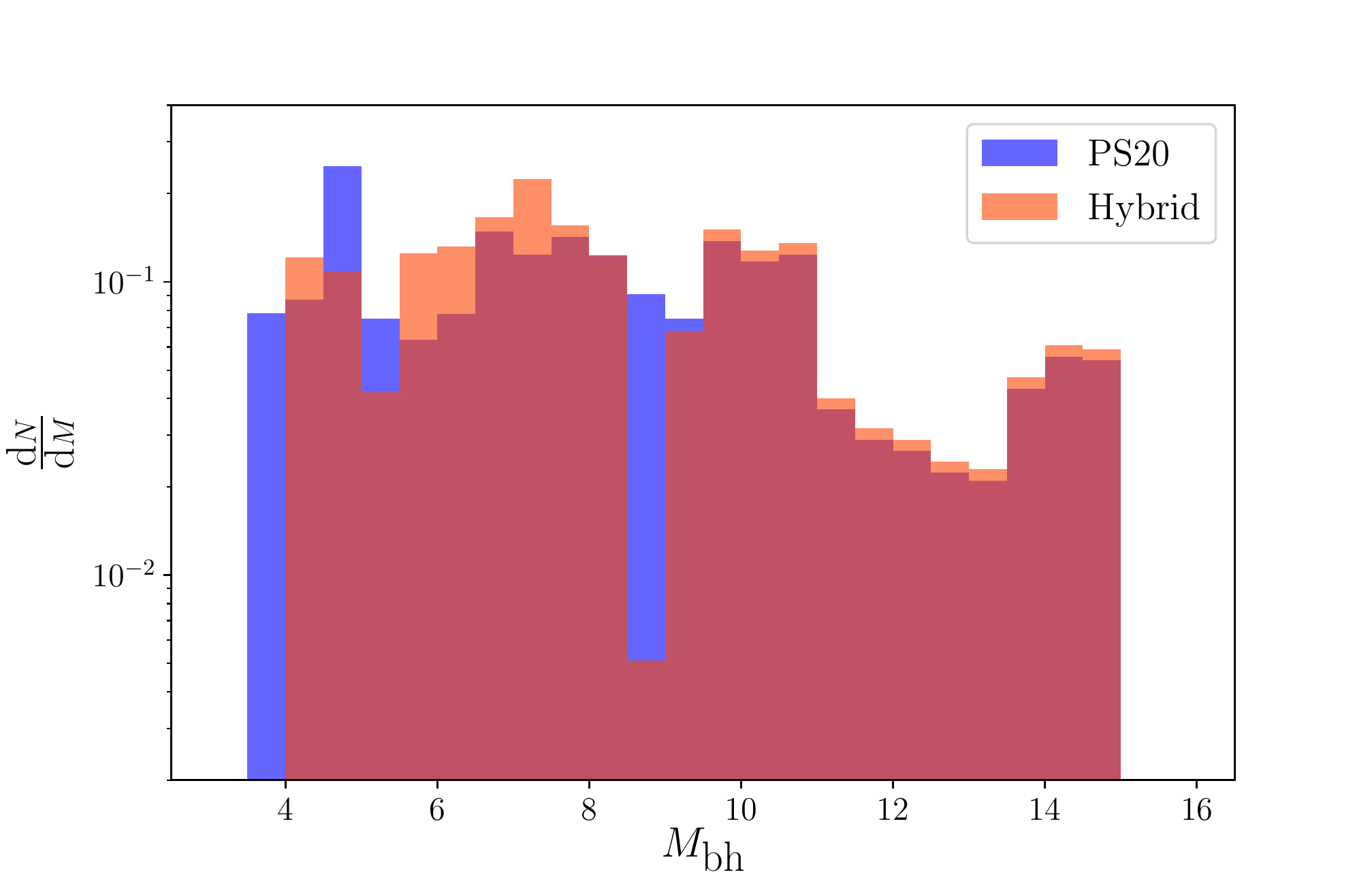}
    \caption{The predicted distributions from binary progenitors using the prescriptions from PS20 (blue) and the hybrid approach (orange) for NSs (top) and BHs (bottom). As with the single star distributions, the overall agreement between the two methods' predicted shapes, frequencies, and mass ranges are good.}
    \lFig{hybin}
\end{figure}

\section{Discussion}\lSect{dis}
One of the limiting factors of comparing these distributions is that the NS and BH mass functions are only somewhat known, making it difficult to compare the predicted distributions to a robust set of observations. There are predictions from theory \citep[e.g.,][]{Pej15,Suk16, Kov17} and limited observations of compact object binaries \citep[e.g.,][]{Oze12, Oze16, Lig20}, which one could qualitatively compare to these or any predicted compact object distributions, noting key similarities like distribution shape and peak value. For example, the PS20 and hybrid NS distributions peak near 1.4 \Msun\ and have a more normal shape whereas the F12 NS distributions peak at 1.2 \Msun\ and drop off. As further observations make the NS and BH mass distributions more robust, comparisons to theoretical predictions for the mass distributions will more concretely show whether any of these BPS prescriptions produce similar distributions.

Still, we now know that a star's explodability depends on its presupernova core structure, which is not considered in the F12 prescription. That is not to say that the PS20 prescription is without its issues. Section 6 of PS20 discusses several key caveats to using bare CO-cores. The main issue is that bare CO-cores are not perfect substitutes for CO-cores embedded in stars. The temperature and pressure gradients are steeper in bare cores than in embedded ones, which leads to non-negligible effects on the core structure. Currently the PS20 models are adequate for predicting bulk trends in populations but for improved accuracy, the study would need to be redone with CO-cores where the effects of a helium shell were also simulated. 

Nevertheless, it is important to examine the implications of using simplified prescriptions for finals fates, remnant types, and remnant masses in BPS, especially if BPS will be used to make predictions for observations, like merger rates from gravitational wave sources. Some recent work is starting to explore this. \citet{Rom20} compare the F12 delayed and rapid prescription and 1D Supernova models from the N20 engine in \citet{Suk16} in the context of progenitors of BH-NS mergers. They find that the N20 models consistently produce up to an order of magnitude more mergers than F12, more consistent with the rates inferred from the first half of the LIGO-Virgo collaboration's third observing run \citep{Abb20a}. Additionally, \citet{Gho21} compares predictions for the distributions of observed chirp masses and mass ratios for detectable LIGO gravitational wave events where remnant masses are determined by four different prescriptions, including the F12 rapid prescription and the default BPASS prescription \citep[see][]{Eld04} in addition to two limiting cases. 

More work like this is warranted, exploring which prescriptions best reflect the observed rates of various gravitational wave sources and the BH and NS mass functions. Recently, \citet{Sch21} used the \citet{Mue16} semi-analytic models to examine the distributions of NSs and BHs from a suite of progenitors stripped by binary interaction. The \citet{Mue16} prescription is similar to the \citet{Ert16, Ert20} criterion adopted by the PS20 prescription \citep[see][]{Suk18}. However, the differences in population explored by this work and \citet{Sch21} make the two difficult to compare. Still, there is no study which compares predictions for something made by every common final fate/remnant mass prescription for the same initial population to observations of said phenomenon. To ensure that BPS codes consistently adopt the most accurate prescription they can, these comparisons must be explored.

There is a hierarchy of accuracy. The ideal approach for predicting final fates or compact object distributions would be to evolve all massive stars in the simulation through to core--collapse and then try to explode the star using a supernova explosion simulation. This is unfeasible for now as the computational expense is quite large. Recent studies of the single and binary progenitors of core--collapse supernovae evolved all the way to core--collapse explore tens to hundreds of stars \citep[e.g.,][]{Suk16, Var21}. Doing this for millions of stars is unrealistic. The computational expense of end-stage modelling could be mitigated in part by probing core structure at core collapse and making predictions for final fate/remnant mass using parameters calibrated from 1D core--collapse supernovae models, as in the two-parameter criterion from \cite{Ert16, Ert20}. 

But BPS codes only evolve stars to central carbon ignition (or slightly later or slightly earlier). The next best way to proceed would be to back out a presupernova core-structure using the PS20 models. Then the user could adopt their preferred method for determining final fate/remnant mass based on the core structure. However, since most BPS codes do not actively evolve their stars, they cannot use the PS20 models. Instead, we recommend adopting the hybrid prescription in rapid BPS codes because it can be implemented easily, while still incorporating the strong dependence final fate has on presupernova core structure. The user can also pick their preferred explosion models to construct the hybrid matrix.

\section{Conclusions}\lSect{Con}
We present a comparison of the compact object distributions predicted by three separate remnant prescriptions, all applied to the same populations of solar metallicity single and binary stars modeled by BPASS. The F12 prescription depends solely on \mco~and the final mass of the star whenever a star’s evolution terminates. For BPS codes, which commonly employ it, that cutoff is at or near central carbon ignition. The PS20 prescription approximates a presupernova structure by interpolating over a grid of CO-cores evolved from central carbon ignition to core--collapse, using values of \mco\ and \xc\ taken from the BPS models at carbon ignition. By applying a core structure-based explosion criterion onto the final core, in this case the \citet{Ert16, Ert20} two-parameter explosion criterion, we predict final fates and remnant masses based on the presupernova core structure. Finally, we adopt a hybrid approach which uses \mhe\ measured at the evolutionary cutoff to interpolate over a table of presupernova \mhe\ and associated remnant masses from detailed evolution and explosion models.  

The differences between the F12 and PS20 prescriptions, outlined in \Sect{single} and \Sect{Bin}, are stark. The F12 prescription predicts a near continuous distribution of remnant masses, including high mass NSs ($M_\mathrm{grav} \approx$ 2.2 \Msun) and high frequency of low mass ($M_\mathrm{bh} \approx$ 2.5 \Msun) BHs (see \Fig{Fry}, \Fig{sing}, \Fig{NSbindist}, and \Fig{BHbindist}). The PS20 prescription shows that core structure limits the possible NS masses and that low-mass, mass gap black holes are much less frequent. NSs produced from higher mass CO-cores, which F12 does not predict, are the direct result of the increase in \xc\ caused by envelope stripping as well as islands of explodability, neither of which F12 is sensitive to. While the BH mass range covered by the two prescriptions' distributions is roughly identical, the distributions themselves differ quite dramatically around BH masses of 10-12.5 \Msun. This is the direct result of envelope stripping driving down the available mass of the star to form a remnant (see \Fig{xc} and \Fig{Fry}). The PS20 models do not include any mass from the envelope in their BH mass calculation, causing no frequency drop in this range. The frequency drop in the PS20 distributions at slightly higher BH masses is due to penalties from the IMF and a small turnover in \mhe\ caused by extreme mass loss.

The differences between the PS20 and hybrid distributions are minimal and largely due to differences in the model physics adopted by PS20 and \citet{Suk16}, from where we draw our models to construct the hybrid matrix (see \Fig{hysin} and \Fig{hybin}). This strong agreement between two different methods which both rely, either explicitly (PS20) or implicitly (hybrid), on presupernova core structure points to the necessity of considering that structure when determining how a star dies and the properties of the remnant it leaves behind. There is a strong correlation between final core structure and final fate. There is not a strong correlation between \mco\ and/or \mfin\ and final fate. 

Qualitatively, the shape and peak of the PS20 and hybrid distributions agree better with other theoretical predictions and limited observations of the NS and BH mass function, compared to F12, the most commonly utilized prescription in BPS studies. Given the integral role played by the presupernova core structure in determining final fates of massive stars and their associated remnant properties, we strongly encourage future BPS calculations to consider and include this structural dependence and utilize modern neutrino-driven explosion results for an improved accuracy of their predictions. As we have shown, the PS20 and hybrid methods are two possible ways to do this, depending on the type of BPS code and the purpose of the study.  

\section{Acknowledgements}
The authors would like to thank E. Stanway and A. Chrimes for their contributions to the BPASS models. JJE acknowledge support from the University of Auckland and also the Royal Society Te Ap\={a}rangi of New Zealand under the Marsden Fund. Support for this work comes from NASA grant 80NSSC19K0597 awarded to M. Pinsonneault. TS was supported by NASA through the NASA Hubble Fellowship grant \#60065868 awarded by the Space Telescope Science Institute, which is operated by the Association of Universities for Research in Astronomy, Inc., for NASA, under contract NAS5-26555.

\section{Data Accessibility}
\lSect{dat}
All models used in the PS20 prescription (\KEPLER\ models, the \MESA\ inlist, and the complete tables of core diagnostics) are available for download at \url{http://doi.org/10.5281/zenodo.3839747}. The specific BPASS data files containing core masses and composition information for this study, as well as a list of the missing models, are available upon request. Standard BPASS output is available at \url{https://bpass.auckland.ac.nz/index.html}.

%%%%%%%%%%%%%%%%%%%%%%%%%%%%%%%%%%%%%%%%%%%%%%%%%%

% Don't change these lines
\bsp	% typesetting comment
\label{lastpage}
\end{document}